\documentclass[fleqn,usenatbib]{mnras}

\usepackage{newtxtext,newtxmath}

\usepackage[T1]{fontenc}

\DeclareRobustCommand{\VAN}[3]{#2}
\let\VANthebibliography\thebibliography
\def\thebibliography{\DeclareRobustCommand{\VAN}[3]{##3}\VANthebibliography}

\providecommand\psj{PSJ}

\usepackage{graphicx}	
\usepackage{amsmath}	
\usepackage{comment}
\usepackage[dvipsnames]{xcolor}
\usepackage[normalem]{ulem}

\title[An \textit{M}-band study of $\beta$ Pic b]{Into the red: an \textit{M}-band study of the chemistry and rotation of $\beta$~Pictoris~b at high spectral resolution\thanks{Based on observations collected at the European Southern Observatory under ESO programme 109.23G2.001}} 

\author[L. T. Parker et al.]{Luke T. Parker,$^{1}$\thanks{E-mail: luke.parker@physics.ox.ac.uk}
Jayne L. Birkby,$^{1}$
Rico Landman,$^{2}$
Joost P. Wardenier,$^{3,4}$
Mitchell E. Young,$^{1,5}$\newauthor
Sophia R. Vaughan,$^{1}$
Lennart van Sluĳs,$^{1,6}$
Matteo Brogi,$^{7,8}$
Vivien Parmentier,$^{9}$ 
Michael R. Line$^{10}$\\
\\
$^{1}$Astrophysics, University of Oxford, Denys Wilkinson Building, Keble Road, Oxford, OX1 3RH, UK\\
$^{2}$Leiden Observatory, Leiden University, Postbus 9513, 2300 RA Leiden, The Netherlands\\
$^{3}$Atmospheric, Oceanic, and Planetary Physics, Clarendon Laboratory, University of Oxford, Oxford, OX1 3PU, UK\\
$^{4}$Département de Physique, Institut Trottier de Recherche sur les Exoplanètes, Université de Montréal, Montréal, Québec, H3T 1J4, Canada\\
$^{5}$Astrophysics Research Centre, Queen’s University Belfast, Belfast, BT7 1NN, United Kingdom\\
$^{6}$Department of Astronomy, University of Michigan, 1085 South University Avenue, Ann Arbor, MI 48109, USA\\
$^{7}$Dipartimento di Fisica, Universit\`a degli Studi di Torino, via Pietro Giuria 1, I-10125, Torino, Italy\\
$^{8}$INAF-Osservatorio Astrofisico di Torino, Via Osservatorio 20, I-10025 Pino Torinese, Italy\\
$^{9}$Université Côte d’Azur, Observatoire de la Côte d’Azur, CNRS, Laboratoire Lagrange, France\\
$^{10}$School of Earth $\&$ Space Exploration, Arizona State University, Tempe AZ 85287, USA\\
}

\date{Accepted 2024 May 10. Received 2024 April 30; in original form 2024 February 29}

\pubyear{2024}

\begin{document}
\label{firstpage}
\pagerange{\pageref{firstpage}--\pageref{lastpage}}
\maketitle

\begin{abstract}
High-resolution cross-correlation spectroscopy (HRCCS) combined with adaptive optics has been enormously successful in advancing our knowledge of exoplanet atmospheres, from chemistry to rotation and atmospheric dynamics. This powerful technique now drives major science cases for ELT instrumentation including METIS/ELT, GMTNIRS/GMT and MICHI/TMT, targeting biosignatures on rocky planets at 3--5~$\mu$m, but remains untested beyond 3.5~$\mu$m where the sky thermal background begins to provide the dominant contribution to the noise. We present 3.51--5.21~$\mu$m \textit{M}-band CRIRES+/VLT observations of the archetypal young directly imaged gas giant $\beta$ Pictoris b, detecting CO absorption at S/N~=~6.6 at 4.73~$\mu$m and H$_2$O at S/N~=~5.7, and thus extending the use of HRCCS into the thermal background noise dominated infrared. Using this novel spectral range to search for more diverse chemistry we report marginal evidence of SiO at S/N~=~4.3, potentially indicative that previously proposed magnesium-silicate clouds in the atmosphere are either patchy, transparent at \textit{M}-band wavelengths, or possibly absent on the planetary hemisphere observed. The molecular detections are rotationally broadened by the spin of $\beta$ Pic b, and we infer a planetary rotation velocity of $v$sin(i)~=~22$\pm$2~km~s$^{-1}$ from the cross-correlation with the H$_2$O model template, consistent with previous \textit{K}-band studies. We discuss the observational challenges posed by the thermal background and telluric contamination in the \textit{M}-band, the custom analysis procedures required to mitigate these issues, and the opportunities to exploit this new infrared window for HRCCS using existing and next-generation instrumentation.

\end{abstract}




\section{Introduction}
\label{sec:Intro}

First demonstrated in the \textit{K}-band by \citet{Snellen2010}, high-resolution cross-correlation spectroscopy (HRCCS) of exoplanets at visible and near-infrared wavelengths has developed into the leading method of ground-based atmospheric characterisation. This powerful technique has substantially advanced our understanding of exoplanet atmospheres, including the detailed understanding of their chemistry and the study of C/O ratios (e.g. \citealt{Brogi2019, Smith2024}) and metallicities (e.g. \citealt{Giacobbe2021}). HRCCS techniques have also probed atmospheric structure including inversion layers (e.g. \citealt{Nugroho2020,Nugroho2021,Pino2020}) and hotspot offsets (e.g. \citealt{Herman2022, Hoeijmakers2022,vanSluijs2023}), alongside investigating atmospheric dynamics such as tidal locking (e.g. \citealt{Brogi2016,Flowers2019,Beltz2021}) and wind speed (e.g. \citealt{Snellen2010,Ehrenreich2020, Wardenier2021, Wardenier2023}). Classical HRCCS, which leverages the Doppler shift of the faint planetary spectrum relative to the quasi-stationary stellar and telluric spectra to disentangle the signals, is typically limited to fast moving, short period planets (a~<~0.05 AU; \citealt{Birkby2018}). This method and has been used to study the hot and ultra-hot Jupiter populations extensively in both transmission and emission (see \citealt{Brogi2021} and references therein).

To study the long-period directly imaged planet population at high spectral resolution, the novel combination of HRCCS techniques with adaptive optics (i.e. high contrast spectroscopy; HCS; \citealt{Sparks2002,Snellen2014,Snellen2015}) is used. This technique, in which the star and planet signals are spatially separated and localised on the detector, has studied the chemistry of young giant planets, including the detection of isotopologues (e.g. \citealt{Zhang2021a}), and has unlocked a new fundamental parameter, the spin rate of freely rotating planets \citep{Snellen2014,Schwarz2016, Bryan2018, Xuan2020, Wang2021a, Landman2023b}. HCS can be performed using slit or fibre fed high-resolution spectrographs (R~$\approx$~100,000), targeting specific spatial regions around the host star; or using an integral field unit (IFU), which permits the application of `molecule mapping' in which the entire image plane of the IFU is analysed \citep{Hoeijmakers2018,Petit2018,WangJi2018,Ruffio2019,Petrus2021}. This latter method is practicable at much lower spectral resolutions (R~$\approx$~2000; \citealt{Landman2023a}), exchanging an additional spatial dimension for reduced spectral resolution. 

\begin{figure}
    \includegraphics[width=\columnwidth]{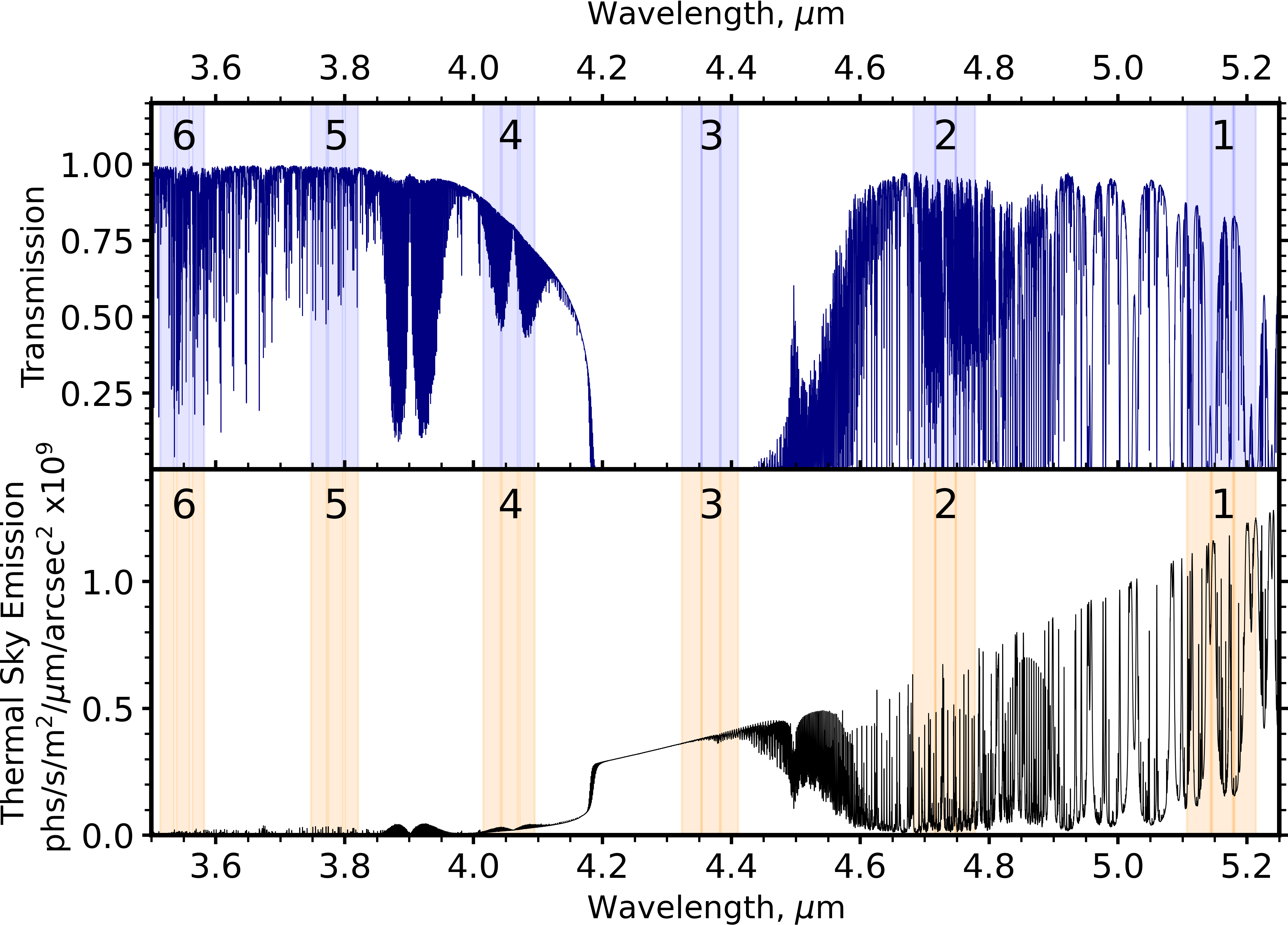}
    \caption{The fractional atmospheric transmission (top) and sky background emission (bottom) in the \textit{M}-band. Highlighted regions denote spectral ranges covered by detector orders in the M4368 grating setting. Note that the 3\textsuperscript{rd} order, at 4.4~$\mu$m, falls at a region of total atmospheric opacity, leaving five usable orders. Spectra generated from the Cerro Paranal Advanced Sky Model \citep{Noll2012,Jones2013}.}
    \label{fig:tellurics}
\end{figure}

The success of HCS, and its unique access to spatially separated targets, has made it a driving science technique for next generation instrumentation including the Mid-infrared ELT Imager and Spectrograph (METIS; \citealt{Brandl2021}) at ESO's Extremely Large Telescope (ELT), which will use HCS to search for biosignatures on the nearest rocky exoplanets in the 3--5~$\mu$m regime, the \textit{L} and \textit{M}-bands \citep{Snellen2015}. Due to the observational challenges of detecting biosignatures with the ELTs in transit \citep{Hardegree-Ullman2023, Currie2023}, the use of HCS is potentially the only prospect of detecting biosignatures in the first decades of ELT operations \citep{Vaughan2024}. However, in order to effectively combine the signal of the many faint spectral lines in the planet’s spectrum, HRCCS techniques have historically been limited to photon noise dominated observations in the visible and near-infrared \citep{Birkby2018}. Unfortunately, in the METIS \textit{M}-band (4--5~$\mu$m) regime, where potential biosignatures (e.g. CO$_2$, CH$_4$, N$_2$O) for rocky planets are expected \citep{Schwieterman2018}, the thermal background starts to become a major noise source. Additionally, all wavelength regions face severe telluric contamination from Earth's atmospheric absorption lines and bright sky emission lines (see Fig.~\ref{fig:tellurics}). This wavelength range poses unique observational challenges, but, to date, HRCCS techniques at R~$\approx$~100,000 have never been tested beyond 3.5~$\mu$m \citep{Webb2020}, and ground-based HCS remains untested and unproven at wavelengths redder than 3.8~$\mu$m \citep{WangJi2018}. In this work, we demonstrate the successful application of these techniques in the 3.51--5.21~$\mu$m regime and push to understand the behaviour of high-resolution spectroscopy in this new thermal noise dominated spectral range. This is a vital step prior to the first light of METIS/ELT and similar \textit{L} and \textit{M}-band high-resolution spectrometers, e.g. the Giant Magellan Telescope Near Infrared Spectrograph (GMTNIRS/GMT; \citealt{Lee2022}) and the Mid-IR Camera, High-disperser \& IFU spectrograph for the Thirty Meter Telescope (MICHI/TMT; \citealt{Packham2018}).

Additionally, opening a new atmospheric window for HRCCS techniques paves the way for the ground-based characterisation of new classes of planets and the study of exotic atmospheric compositions. Of particular interest in the \textit{M}-band are the so-called `lava worlds', rocky super-Earths with ultra-short orbital periods and dayside temperatures in excess of 2000 K, predicted to host hemispheric magma oceans, analogous to the early-Earth \citep{Leger2011, Chao2021,Lichtenberg2023}. Silicate species vaporised from the surface, such as SiO, SiO$_2$, and MgO are predicted to form thin rock vapour atmospheres \citep{Schaefer2009,Miguel2011, Kite2016,Zilinskas2022}, alongside volatile species outgassed from the interior including H$_2$O, CO, and CO$_2$ \citep{Zilinskas2023,Piette2023}, all of which show strong opacities and spectral features in the \textit{M}-band. Distinct spectroscopic features, paired with the favourable planet-star contrast ratio for these ultra-hot planets, mean lava worlds may be highly observable in the \textit{M}-band.

Using the newly upgraded Cryogenic High-Resolution In-frared Echelle Spectrograph (CRIRES+; \citealt{Dorn2014,Dorn2023}) at the VLT we observe the young, widely-separated planet $\beta$ Pictoris b in the \textit{M}-band. $\beta$~Pic b, with a bright host star (\textit{M$^{\prime}$}~=~3.458$\pm$0.009~mag; \citealt{Stolker2020}), favourable \textit{M}-band planet-star contrast ratio ($\sim$$10^{-4}$), and previous high-resolution detections of CO and H$_2$O \citep{Snellen2014, Hoeijmakers2018,Landman2023b}, is a compelling target for verifying the use of HRCCS in the \textit{M}-band. In Section \ref{sec:Beta Pic System} we outline the architecture and previous study of the $\beta$~Pic system, and in Section \ref{sec:Observations} we detail our CRIRES+ observations. Section \ref{sec:Methods} discusses our custom analysis procedure required to process high spectral resolution \textit{M}-band observations. In Sections \ref{sec:Results} and \ref{sec:Discussion} we present our results and discuss their implications for future studies. We conclude in Section \ref{sec:Conclusions}. 

\section{The \texorpdfstring{$\beta$}{Beta} Pictoris System}
\label{sec:Beta Pic System}

$\beta$ Pictoris is a young (23 Myr) A5V star \citep{Lagrange2019}, which shows variability in the form of $\delta$ Scuti pulsations \citep{Koen2003}. The system famously hosts an edge-on circumstellar debris disc, extending beyond 1000 AU \citep{Larwood2001}, which has been extensively studied due to the proximity (19.44$\pm$0.05 pc; \citealt{vanLeeuwen2007}) and brightness (\textit{K} = 3.48 mag; \citealt{Ducati2002}) of the star. The disc is predicted to host belts of planetesimals \citep{Thebault2001}, resulting in in-falling comets \citep{Kiefer2014, Zieba2019}. These structures, alongside a warped inner disc component, led to suggestions of a planetary-mass companion on an eccentric orbit at $\sim$10~AU \citep{Scholl1993, Roques1994, Lazzaro1994, Mouillet1997, Augereau2001}. The predicted planet, $\beta$ Pic b, was discovered using direct imaging with NACO/VLT \citep{Lagrange2009, Lagrange2010}. The orbit of $\beta$ Pic b is closely aligned to the line of sight from Earth \citep{Wang2016} with a semi-major axis 9.8$\pm$0.4~AU and non-zero eccentricity 0.09$\pm$0.1 \citep{Lagrange2020}. Despite the well aligned orbital orientation the planet is non-transiting; a predicted transit of the planet's Hill sphere between April 2017 and January 2018 \citep{LecavelierdesEtangs2016, Wang2016} revealed no photometric transit signals \citep{Kenworthy2017,vanSluijs2019,Kenworthy2021}. Numerous studies using high contrast imaging brightness-mass relationships \citep{Lagrange2010}, interferometric measurements \citep{GRAVITY2020}, astrometry \citep{Kervella2019,Lacour2021}, and including radial velocity measurements \citep{Snellen2014, Dupuy2019, Nielsen2020, Lagrange2020, Landman2023b}, have been carried out to constrain the mass, radius, and precise orbital solution of $\beta$ Pic b. Our selected properties of the system are given in Table~\ref{tab:parameters_table}.

\setlength{\tabcolsep}{3pt}
\begin{table}
	\centering
	\caption{The stellar and planetary parameters of the $\beta$ Pictoris system.\label{tab:parameters_table}}
	\begin{tabular}{ccl} 
		\hline
		Parameter & Value & Reference\\
		\hline
            \hline
             & \textbf{          $\beta$ Pic A: stellar parameters} & \\
            \hline
            RA & 05$^{\text{h}}$ ~47$^{\text{m}}$~17$^{\text{s}}$.0876901 & \citet{vanLeeuwen2007}\\
            Dec & -51\degr~03\arcmin~59.441135\arcsec & \citet{vanLeeuwen2007}\\
            \textit{M$^{\prime}$} & 3.458$\pm$0.009~mag& \citet{Stolker2020} \\ 
		R$_{\star}$ & 1.544~R$_{\sun}$ & \citet{Stassun2019} \\ 
		M$_{\star}$ & 1.795$\pm$0.027~M$_{\sun}$ & \citet{Feng2022}\\ 
		T$_{\text{eff}}$ & 7890 $^{+13}_{-17}$~K & \citet{Swastik2021}\\ 
            Spectral type& A6V & \citet{Bonnefoy2013}\\ 
            $[$Fe~/~H$]$ & -0.21 $^{+0.03}_{-0.02}$ & \citet{Swastik2021} \\
            Age & 23~Myr & \citet{Lagrange2019}\\
            Distance & 19.44$\pm$0.05~pc & \citet{Bonnefoy2013} \\
            $v_{\text{sys}}$ & 20.0$\pm$0.7~km~s$^{-1}$ & \citet{Gontcharov2006}\\ 
		\hline
             & \textbf{             $\beta$ Pic b: planetary parameters} & \\
            \hline
            M$_{\text{p, Dynamical}}$ & 9.3$^{+2.6}_{-2.5}$~M$_J$ & \citet{Brandt2021}\\ 
		M$_{\text{p, Photometric}}$ & 12.7$\pm$0.3~M$_J$ & \citet{Morzinski2015}\\
            R$_{\text{p}}$ & 1.45$\pm$0.02~R$_J$ & \citet{Morzinski2015}\\
		T$_{\text{p}}$ & 1708$\pm$23~K & \citet{Morzinski2015}\\
            \textit{M$^{\prime}$} &  10.96$\pm$0.13~mag& \citet{Currie2013}\\
            P$_{\text{orb}}$ & 23.6 $^{+0.3}_{-0.2}$ years & \citet{Feng2022}\\
            P$_{\text{rot}}$ & 8.1$\pm$1.0~h & \citet{Snellen2014}\\ 
            \hline
            \hline
	\end{tabular}
\end{table}

The archetypal young self-luminous giant planet, $\beta$ Pic b has been well studied spectroscopically, including two high-resolution HCS studies in the \textit{K}-band (\citealt{Snellen2014}, the original CRIRES; \citealt{Landman2023b}, CRIRES+) and the use of medium-resolution `Molecule Mapping' (\citealt{Hoeijmakers2018,Kiefer2024}, SINFONI; \citealt{Worthen2024}, MIRI). The current state of the art at low-resolution is the GRAVITY/VLT \textit{K}-band spectroscopy \citep{GRAVITY2020} combined with GPI/Gemini spectroscopy \citep{Chilcote2017}. Together, these studies have revealed an atmosphere containing molecular CO and H$_2$O, an effective temperature T$_{\text{eff}}$~$\approx$~1700~K with low surface gravity \citep{Hoeijmakers2018,GRAVITY2020,Landman2023b}, and the first measurement of an exoplanetary rotation period: 8.1$\pm$1.0 hours \citep{Snellen2014}, recently remeasured to 8.7$\pm$0.8 hours \citep{Landman2023b}. These previous studies provide a benchmark for our observations in the \textit{M}-band. An additional, closer in planet, $\beta$ Pic c, has been detected using HARPS radial velocity measurements \citep{Lagrange2019} and astrometrically confirmed with GRAVITY/VLT \citep{Nowak2020}. At the time of observation $\beta$~Pic~c is separated from the host star by only 0.011\arcsec, $\approx$~2 CRIRES+ spatial pixels \citep{Wang2021b}.

\label{Beta}
\section{Observations}
\label{sec:Observations}

We observe the $\beta$ Pictoris system with the recently upgraded CRIRES+, mounted at the Nasmyth B focus of the VLT UT3 (Melipal). Our observations, totalling 2.4 hours including overheads, are taken in two observing blocks (OBs) on 5th and 11th April 2022 (Program ID: 109.23G2.001, PI: Birkby). CRIRES+ operates as a cross-dispersed slit spectrograph, with a slit length of 10\arcsec. The slit width of 0.2\arcsec is chosen to achieve the maximum spectral resolving power offered by CRIRES+ in the \textit{M}-band, nominally R~$\sim$~92~000 for observations with uniform illumination of the slit. The slit is oriented along the star-planet axis to include both the host star and the target planet $\beta$ Pic b, with a position angle PA~=~31.4\degr. Each observing block consists of 102 science exposures of 20s, taken in the classical ABBA nod pattern during which the telescope is nodded 6\arcsec along the slit, to allow accurate background subtraction, vital for the background-dominated \textit{M}-band observations. We chose the M4368 grating setting, which provides non-continuous wavelength coverage across the range 3.51--5.21~$\mu$m over six orders, covering regions of the \textit{L} band, and spanning the \textit{M}-band, approaching the CRIRES+ detector cut off at 5.3~$\mu$m (see Fig.~\ref{fig:tellurics}). This grating setting targets wavelength regions with lower telluric contamination, higher transparency, and which contain strong spectral features of target molecules in the planetary atmosphere. Our observations were taken in good turbulence conditions, with average DIMM measured seeing of 0.637\arcsec and 0.630\arcsec for OB1 and OB2, respectively (see Table~\ref{tab:obs_table}). The further use of the Multi-Application Curvature Adaptive Optics system (MACAO; \citealt{Paufique2004}) results in a suppression of the starlight at the position of the planet, measured to reach a maximum suppression factor of $\sim$100. This additional suppression of the stellar spectrum is vital for the detection of the planetary companion $\beta$ Pic b, located at the time of observation at a separation of 0.506\arcsec \citep{Wang2021b}, nine CRIRES+ spatial pixels. We observe that the stellar signal is significantly weakened in all raw frames at the B nod position for OB2. The source of this effect is unclear, but we suggest a likely cause is pointing errors in the nodding throw, resulting in a misalignment between the slit and the target. The average S/N of the B nod frames are dramatically reduced, with an S/N three times lower than the average S/N of all other exposures; consequently, we exclude the OB2 B nod positions from our analysis. The A nod positions in OB2 are unaffected. Additionally, the third detector order lies in an opaque atmospheric region due to the 4.4~$\mu$m telluric CO$_2$ band (Fig.~\ref{fig:tellurics}) and is likewise excluded from our analysis for both OBs.

\setlength{\tabcolsep}{6pt}
\begin{table}
	\centering
	\caption{Details of our CRIRES+ observations of $\beta$ Pic b. The average precipitable water vapour (PWV) and DIMM measured seeing are recorded from conditions at Paranal at the time of observation.\label{obs_table}}
	\label{tab:obs_table}
	\begin{tabular}{lcc} 
		\hline
             & OB1 & OB2\\
		\hline
            \hline
            Date & 05/04/2022 & 11/04/2022 \\
            Av.~PWV & 1.76~mm & 1.31~mm \\
            Av.~Seeing & 0.637\arcsec & 0.630\arcsec \\
            Av.~Airmass & 1.35 & 1.57 \\
            N$_{\text{exp}}$ & 102 & 102 \\
            DIT & 20~s & 20~s \\
            Slit Width & 0.2\arcsec & 0.2\arcsec\\
            Position Angle & 31.374\degr & 31.378\degr\\
            Planet Sep. & 0.5055\arcsec & 0.5065\arcsec\\
            Grating Setting & M4368 & M4368\\
            Resolution & $\approx$~100,000 & $\approx$~100,000\\
            $\lambda$ Coverage & 3.51--5.21~$\mu$m& 3.51--5.21~$\mu$m\\

		\hline
  \hline
	\end{tabular}
\end{table}

\begin{figure}
    \includegraphics[width=\columnwidth]{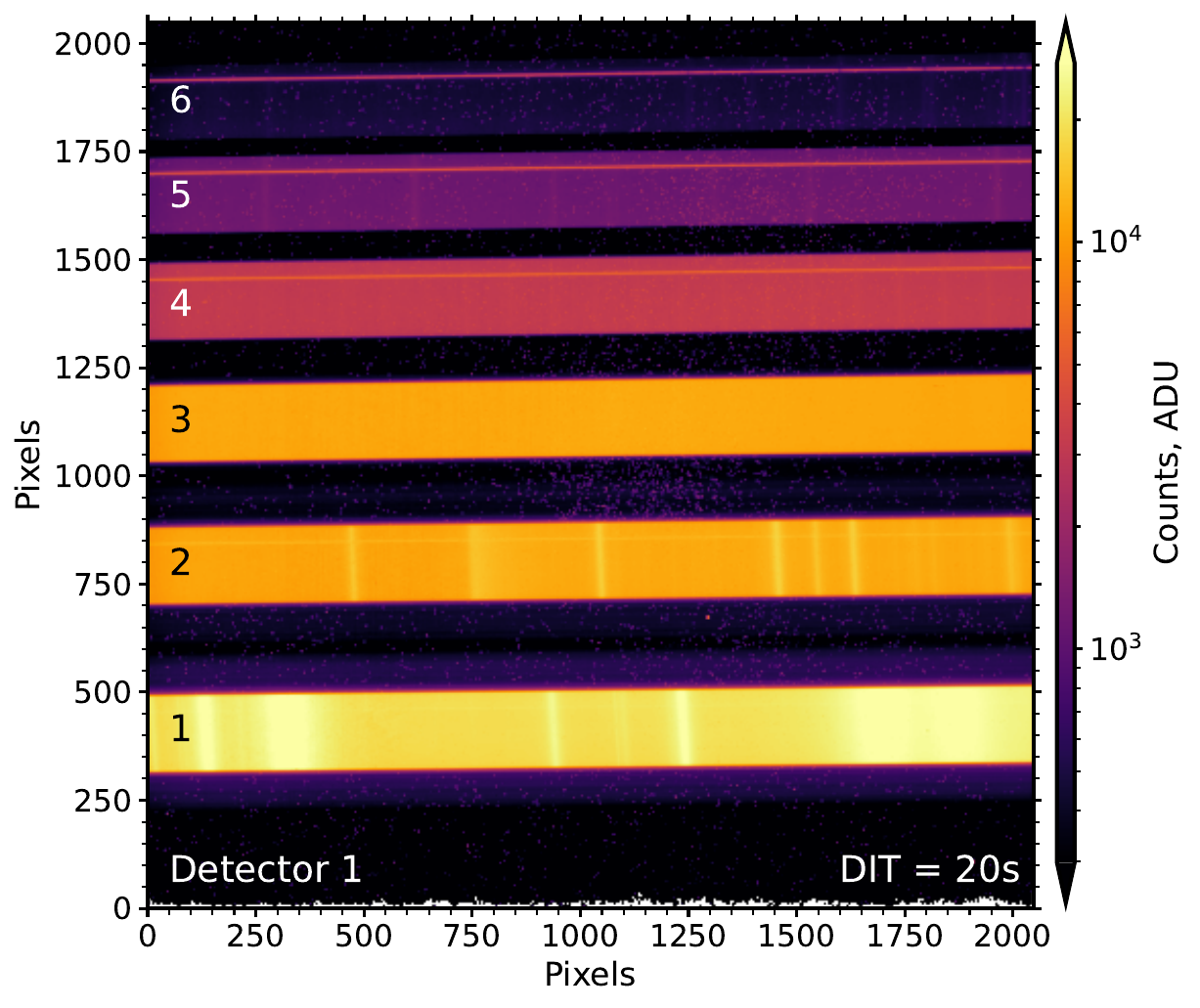}
    \caption{One of the three raw CRIRES+ detector frames for a single 20 second exposure in the \textit{M}-band using the M4368 grating setting. Note the dramatic increase in sky background emission as the orders progress from the 3.5~$\mu$m order, dispersed across the top of the detector; to the reddest order at 5.2~$\mu$m, dispersed at the bottom of the detector.}
    \label{fig:raw_frame}
\end{figure}

\section{Methods}
\label{sec:Methods}

Our analysis procedure is initially based on \textit{pycrires}\footnote{\url{https://pypi.org/project/pycrires/}} \citep{pycrires}, an open-source python wrapper for the EsoRex cr2res routines. Alongside basic calibrations, \textit{pycrires} also provides custom functions specifically developed for HCS observations of sub-stellar companions \citep{Landman2023b}. Our analysis approach is to extract a spectrum from the detector at each position along the slit and then clean the data to remove telluric and stellar contributions. The planet signal, which is present only at the planet spatial position, is buried in the noise, and is extracted through cross-correlation with model planet spectra. However, due to the unique challenges posed by \textit{M}-band data, substantial modifications to many of the basic calibration routines of \textit{pycrires} are required, and novel reductions are essential for many stages of data post-processing. In this section, we will highlight key areas of the traditional HRCCS analysis procedure which are ill-equipped to deal with the \textit{M}-band regime, demonstrate our approaches to mitigating these issues, and suggest areas for investigation in future works.

\subsection{Basic calibrations}
\label{sec:Basic calibrations}

Initial image processing is performed by the pipeline with no modification. The pipeline uses the AB nodding pairs to perform background subtraction, crucial to the success of HCS in the thermal background dominated \textit{M}-band. At both the A and B nod positions a 4\arcsec wide trace is extracted from each background subtracted frame, at the 0.059~\arcsec/pix spatial pixel scale of the CRIRES+ detectors. Each nod position and observing block is analysed independently and only combined after the cross-correlation stage. The spectral orders on the detector are curved due to their projection on the CRIRES+ detector array (Fig.~\ref{fig:raw_frame}). Tracing this curvature is typically performed by the default ESO pipeline by tracking the curved orders from the master flat field. We find a slight misalignment between the flatfield-based pipeline tracing and the curvature of the orders on the science frames, and therefore develop a custom algorithm to trace the spectral orders directly from the science frames. The high thermal background emission in the \textit{M}-band can be leveraged to trace the order boundary of the detector orders, which we model with a second order polynomial fit.

\begin{figure}
    \includegraphics[width=\columnwidth]{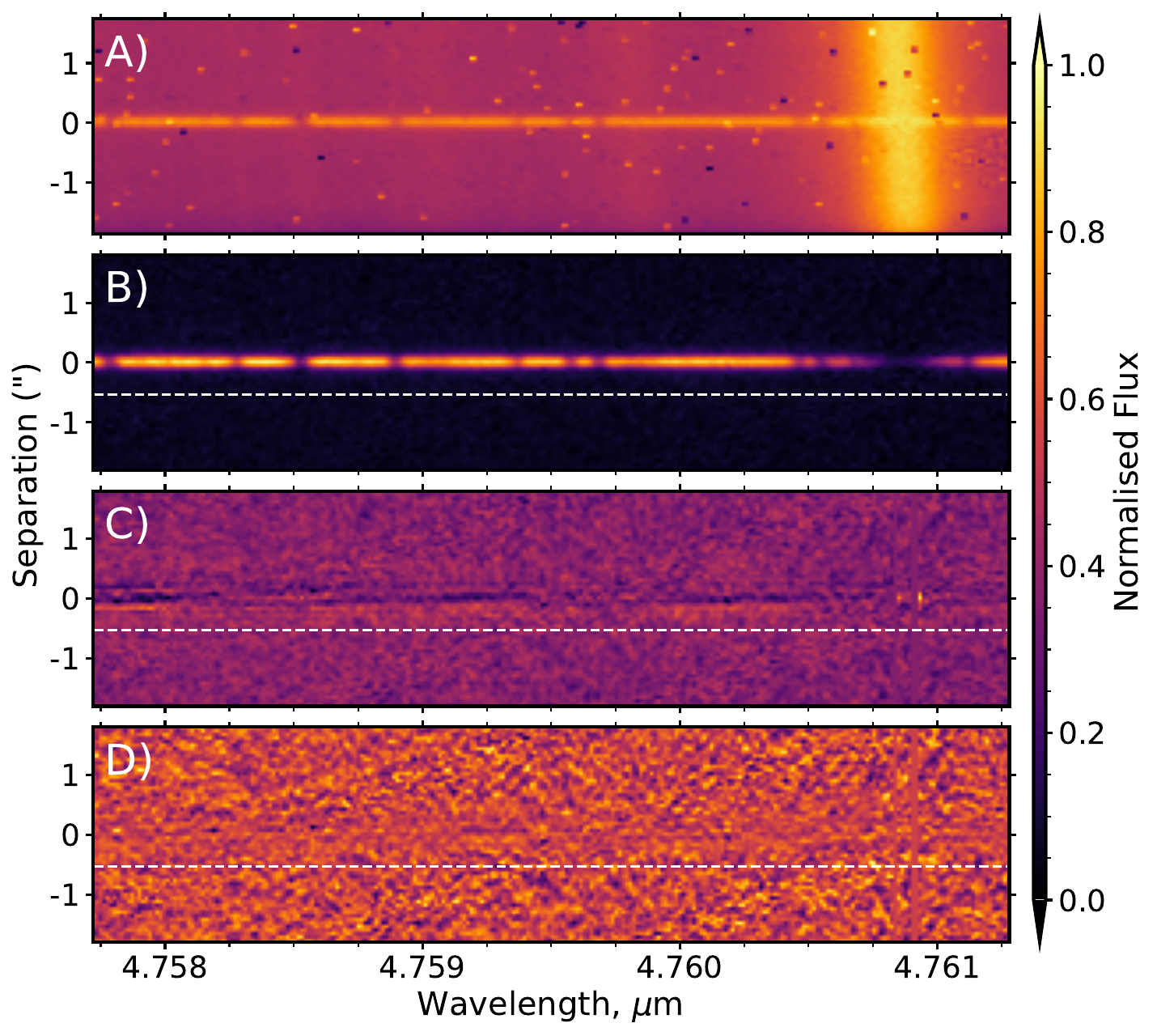}
    \caption{The processing stages of our analysis for a sample wavelength region of the second order. Panel A shows the raw data, demonstrating artefacts including bad pixels, tilted sky emission lines, and high thermal background. These systematics are corrected through the AB nodding procedure, slit extraction, and bad pixel correction. The extracted, bad pixel corrected, and wavelength calibrated 2D spectra, B, undergo masking and subtraction of the stellar trace to produce C. Residuals in the stellar trace resulting from this process, alongside telluric residuals, are removed by normalisation and the iterative removal of PCA components. The resulting cleaned spectra, D, are subsequently cross-correlated with planetary models. The expected location of the planetary companion $\beta$ Pic b is marked in white. Note that the diffraction effects discussed in Section \ref{sec:Bad Pixels} are not visible in this plot.}
    \label{fig:data_processing}
\end{figure}

A further important consideration is the slit image curvature, i.e. the imprinted curvature of static spectral features across the spatial extent of each order, resulting from slit diffraction effects (see slight tilt in vertical sky emission line in Fig.~\ref{fig:data_processing}, panel A). A robust measurement of slit curvature is crucial to the success of spatially resolved HCS, as each spatial pixel is required to be aligned following extraction. Since wavelength calibration is only performed on the central stellar trace (see Section \ref{sec:Wave Cal}), it is vital that the pixels away from the trace, where the planet signal is found, are aligned to the stellar signal pixel-space, as they do not have an independently calibrated wavelength solution. In the CRIRES+ \textit{YJHK} grating settings this can be accurately calibrated using the Fabry Perot Etalon (FPE) system, which provides a comb of reference lines, evenly spaced in frequency. The FPE, however, is designed for use at wavelengths $<$~2.5~$\mu$m and is therefore unsuitable for observations with the M4368 setting. 

Instead, we measure the slit image curvature using a custom routine to track the first order tilt of the sky emission lines in the raw science frames, as suggested in \citet{Holmberg2022}. The sky emission lines are relatively sparse and, due to the high thermal background in the raw images, have a low S/N even in the reddest orders, but are approximately evenly distributed across the detector orders. To track the curvature we therefore stack all raw science frames for each OB to optimise the S/N. The sky emission lines are fitted with a Gaussian line profile for each spatial position, and the shift of the peak across the spatial direction of each order is fit with a first order polynomial. The slit tilt is assumed to remain constant across the detector dispersion direction. Improvements to this algorithm would consider the curvature of the slit image by introducing a second order fit and allow the slit curvature to vary across the detector dispersion direction, but due to the sparse, low S/N, sky emission lines these additions were found to significantly impair the accuracy of the fit. Following extraction, any residual curvature of the spectral orders on the detector is imprinted on the optimally extracted 2D spectra (see \citealt{Schwarz2016}). To rectify these remaining residuals, a Gaussian profile is fitted to the spatial direction of the stellar trace and the data is interpolated to centre the stellar signal, and therefore the planet signal, into a single row across the order.

\subsection{Bad pixels and diffraction artefacts}
\label{sec:Bad Pixels}

The bad pixel correction routine in the ESO CRIRES+ pipeline is highly adaptable but, due to the large dynamic range of the background thermal emission in the \textit{M}-band frames, it is unable to detect all bad pixels, necessitating a custom bad pixel identification and correction procedure. Following spectral extraction we detect remaining outliers on the scale of a single pixel through application of a Laplacian of the Gaussian algorithm, or `Blob Detection Algorithm' (e.g. \citealt{Kong2013,vanSluijs2023}). Pixels that deviate by $>$~5$\sigma$, representing significant outliers from cosmic rays or bad detector pixels, are flagged and corrected through interpolation over the nearest good pixels in the dispersion direction. Bad pixel correction is performed iteratively until no further bad pixels are identified, up to a maximum of 10 iterations, beyond which there is no significant improvement in the number of bad pixels identified. 

Due to the excellent performance of the CRIRES+ MACAO system, and the increased Strehl ratio in the \textit{M}-band, the stellar PSF is sufficiently narrow that it does not evenly fill the slit. This effect, known as super-resolution, produces observations with spectral resolution dominated by the PSF width, rather than the slit width. In this regime we anticipate wavelength shifts on the order of one pixel between reduced spectra from the A and B nodding positions, which is mitigated by independent calibration of each nod. Cross-correlation of model telluric templates with the data following masking of the continuum favours a spectral resolution of R~$\approx$~100,000 for each observing block. An additional consequence of the high Strehl ratio is the presence of diffraction effects from the PSF of the star and the slit optics, most prominently seen in a faint `secondary trace' located at $\approx$ 1.5$\lambda/D$ arcsec from the stellar signal. The spatial location of this secondary trace is wavelength dependent, eliminating the possibility of an astrophysical origin, and we interpret this to be the first subsidiary maximum of the stellar PSF. No further action is required to process these diffraction artefacts, as they are removed alongside the stellar trace, and do not show signals in the CCFs. Nonetheless, we highlight these signals as they can visually mimic the signature of an astrophysical companion.

\subsection{Wavelength calibration}
\label{sec:Wave Cal}

HCS is highly sensitive to the location and shape of the planetary molecular lines, and thus requires precise wavelength calibration. CRIRES+ is equipped with a range of stable wavelength reference sources including a UNe lamp, a short gas cell with multi-species gas fillings (NH$_3$, $^{13}$CH$_4$, C$_2$H$_2$), alongside the Fabry-Perot etalon system \citep{Dorn2023}. The only wavelength reference covering the \textit{L} and \textit{M}-bands, however, is the N$_2$O gas cell, which has sparse line coverage for the wavelengths covered by the M4368 grating. We instead develop a custom wavelength calibration procedure using the high-density of telluric lines in our data. \textit{Tellfit} \citep{Gullikson2014} line-by-line radiative transfer modelling is used to generate a high-resolution telluric reference model for each frame, using the recorded observing conditions of each exposure. Comparing this synthetic telluric reference to the telluric lines in the stellar trace, we make a first guess to the wavelength solution by fitting a second order (quadratic) polynomial to the default M4368 wavelength solution. These parameters are used as priors to generate a grid of wavelength solutions, which are subsequently cross-correlated with each detector order for each exposure. The model which maximises the cross-correlation for each order is considered the best match to the wavelength solution and adopted as the prior for the next grid of models, with this process iterating with a refining grid size until a stable solution is reached. We estimate a minimum wavelength calibration accuracy of $\lesssim$~$\pm$0.05~nm, with a theoretical maximum accuracy of the order $\pm$10$^{-4}$ nm. The \textit{Tellfit} synthetic model is limited to wavelengths below 5~$\mu$m, and therefore the telluric reference model for the calibration of the reddest order (5.13--5.21~$\mu$m), is generated by \textit{Molecfit} \citep{Smette2015}. Calibration is also tested with \textit{Molecfit} providing the stable telluric reference for all orders, but the generated models show a sub-optimal match to our data at shorter wavelengths. The stellar spectrum shows only a few shallow lines in our orders, and therefore does not significantly contaminate the calibration. 

Due to the rapidly declining S/N ratio at spatial separations away from the stellar trace, this calibration can only be accurately performed for the central rows containing the stellar spectrum, and the solution must be extended to all rows of the detector. We find, however, no evidence of the solution diverging over the central rows of the detector and are therefore reasonably confident to extrapolate the solution to all spatial positions. Since we have modelled the slit curvature, the extracted signal at the planet location is aligned in wavelength to the stellar trace where our wavelength solution is defined.

\subsection{Post-processing}
\label{sec:Post-Processing}

The planetary signal, which we expect to be located nine pixels from the central stellar trace, is at this stage buried beneath the background noise, the stellar spectrum, and telluric contamination (see dashed white line, Fig.~\ref{fig:data_processing}, panel B). Both the continuum emission and the spectral lines in the \textit{M}-band are dominated by the Earth’s atmosphere, with very few and comparatively weak lines from the star. Due to the high telluric contamination across all orders in the \textit{M}-band, we combine the methods of previous HCS studies \citep{Snellen2014, Schwarz2016, Hoeijmakers2018, Landman2023b} to remove the telluric contamination, stellar spectral lines, and continuum simultaneously. First, the deepest telluric lines are masked. For each order, we fit the continuum emission using a Gaussian smoothing of the spectra. Columns that deviate by greater than one standard deviation from the continuum are masked, removing the cores of the deepest telluric lines. Additionally, we stipulate that gaps of less than three pixels between masked regions are also masked, to avoid spurious columns disproportionately biasing the subsequent systematic removal routine and cross-correlations. Remaining outliers at the pixel level are flagged and interpolated over, removing any bad pixels that persist following our bad pixel correction routines.

To remove the stellar trace, and subtract the telluric contamination from each spatial position, we first follow the methods of \citep{Hoeijmakers2018}. For each order and each exposure, a master spectrum is extracted from the central rows, containing the flux from the star and the telluric contamination. The spatial information provided by HCS permits the construction of a master spectrum for each frame, capturing the time dependence of the telluric features, and frame specific line distortions from the pulsating star. The CRIRES+ data appears to suffer from a sinusoidal systematic in the extracted spectra, with variable period and phase across spatial positions in the extracted spectra. The master spectrum must therefore be adjusted to remove this periodicity. At each spatial position, the modulation of the sinusoid is determined by dividing the spectrum of each row by the master spectrum, followed by convolution with a Gaussian with a 1$\sigma$ width of 501 pixels. This modulation is applied to the master spectrum for each row, which is additionally corrected by the use of a singular value composition (SVD) kernel of width 50 pixels to determine the line spread function for each spatial position \citep{Snellen2014, Schwarz2016}, and thus the phase offset of the sinusoidal systematic. Finally, we subtract the master spectrum from the local spectrum, suppressing the effect of telluric and stellar lines across all spatial positions. This process is challenging to perform in the \textit{M}-band as the strong telluric contamination adds low-frequency structure to the spectra, complicating the determination of the modulation and phase shift of the sinusoidal systematic. Away from the stellar trace the thermal background noise rapidly becomes dominant over the sinusoidal systematic structure.

Following the removal of the stellar and telluric signals significant residuals remain, including low-frequency periodicity from imperfect correction of the sinusoidal systematic, most prominent in the stellar trace of redder orders (see Fig.~\ref{fig:data_processing}, panel C). To remove these components, alongside remaining correlated noise and telluric contamination, we employ Principal Component Analysis (PCA; \citealt{Murtagh1987,Press1992}). Each successive component of the PCA removes systematic trends in the spatial dimension of the noisy \textit{M}-band data, but the `trend' of the planetary signal is also continually eroded. Careful optimisation of the number of components removed is therefore vital to preserve the planetary signal. As we remove PCA components we record the variance of the resulting data, and choose the optimum number of removed components to be the point at which the reduction in variance plateaus (e.g. \textcolor{blue}{Spring et al. in prep}), which is determined to be eight PCA components in our data (see Appendix~\ref{sec:App}). We hold the number of PCA iterations fixed for both OBs and all detector orders, in order to avoid the optimisation of order specific systematic effects \citep{Cabot2019, Spring2022, Cheverall2023}. Adopting the approach of \citet{Landman2023b} we exclude the row known to contain the planetary signal (alongside the adjacent $\pm$1 rows) from the calculation of PCA components, to avoid the planetary signal influencing the modelling of the systematic trends.

\subsection{High-resolution models}
\label{sec:Models}

\begin{figure}
    \includegraphics[width=1\columnwidth]{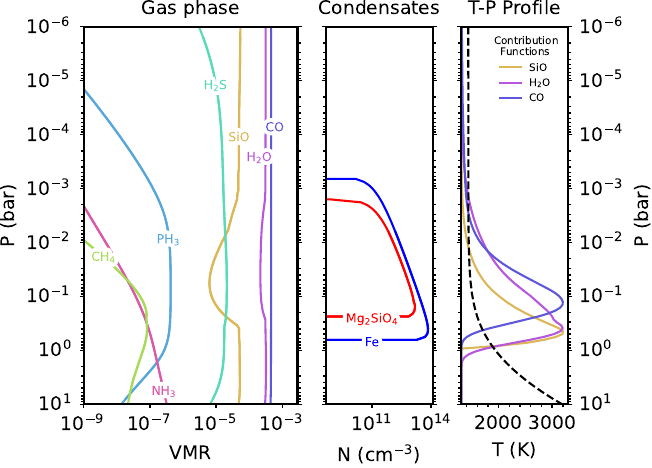}
    \caption{Left: The predicted molecular abundances in the atmosphere of $\beta$~Pic~b in volume mixing ratio (VMR), generated from the \textit{FastChem Cond} equilibrium chemistry code \citep{Stock2018, Kitzmann2024}, including condensation. Centre: The predicted number density (N) of condensable species. Right: The contribution curves for each of the key targeted molecules, demonstrating the atmospheric levels the \textit{M}-band data is sensitive to, and the parameterised T-P profile used in this work (black dashed curve).}
    \label{fig:fastchem}
\end{figure}

In order to extract the comparatively weak planetary signal from the processed data, we require high-resolution spectra for molecular species of interest, derived from atmospheric models. These model spectra are used as cross-correlation templates (see Section \ref{sec:Cross-correlation}) in order to detect the presence of each molecule in the planetary atmosphere. Template emission spectra, informed by the \textit{FastChem Cond} equilibrium chemistry model (\citealt{Stock2018, Kitzmann2024}; see Fig.~\ref{fig:fastchem}), are produced using the open-source radiative transfer code \textit{petitRADTRANS} \citep{Molliere2019}, accounting for collisionally induced absorption of H$_2$-H$_2$ and H$_2$-He, and Rayleigh scattering from H$_2$ and He. The metallicity of the host star, [Fe/H]~=~-0.21$^{+0.03}_{-0.02}$ \citep{Swastik2021}, is used and the spectra are convolved to the observed CRIRES+ instrument resolution, R~=~100,000. As a widely separated self-luminous planet, the atmospheric structure of $\beta$ Pic b is dominated by internal heating. We adopt a parameterised analytical T-P profile of the form:

\begin{equation}
T^4 = \frac{3T_{\text{int}}^4}{4} \left[\frac{2}{3} + \frac{P\kappa_{\text{IR}}}{g}  \right]
\end{equation}

where T$_{\text{int}}$ is the internal temperature, $g$ the gravity, P the pressure, and $\kappa_{\text{IR}}$ the atmospheric opacity at infrared wavelengths. We select values of T$_{\text{int}}$~=~1700~K, log(g)~=~3.5, and $\kappa_{\text{IR}}$~=~0.005~cm$^2$~g$^{-1}$ to best represent $\beta$ Pic b \citep{Morzinski2015,Hoeijmakers2018,Guillot2010}. This profile, representing a non-irradiated, grey, radiative atmosphere, is equivalent to the standard \citet{Guillot2010} profile in the case of minimal irradiation (i.e. T$_{\text{irr}}$ $\ll$ T$_{\text{int}}$). A purely radiative temperature structure is a good approximation for the pressure levels that we are sensitive to, and the \textit{M}-band observations probe a maximum depth of $\sim$1 bar (see Fig.\ref{fig:fastchem}), while the convective region of $\beta$ Pic b is predicted to lie deeper in the atmosphere, at pressures $\leq$~10~bar \citep{Baudino2015}.

Following the above, we produce and cross-correlate line-by-line model emission spectra for the dominant, and most readily detectable, molecular species in equilibrium; H$_2$O (main isotopologues; \citealt{Rothman2010}) and CO (all isotopologues; \citealt{Rothman2010}), alongside the trace molecular species SiO \citep{Barton2013}, H$_2$S \citep{Rothman2013}, PH$_3$ \citep{Sousa-Silva2015}, CH$_4$ \citep{Yurchenko2014}, and NH$_3$ \citep{Yurchenko2011}. The models adopted for CO, H$_2$O, and SiO are shown in Fig.~\ref{fig:models}, while those for H$_2$S, PH$_3$, CH$_4$, and NH$_3$ are shown in Appendix~\ref{sec:App}. Note that while the models used for cross-correlation in this work reflect the molecular abundances derived from the predicted condensation of magnesium-silicate species, they do not include cloud opacities. Prior to cross-correlation (see Section~\ref{sec:Cross-correlation}), the template emission spectra are continuum subtracted, and convolved with a Gaussian with a 1$\sigma$ width of 501 pixels, replicating the mild high-pass filtering applied to the data. This broad filter does not appreciably impact the spectral lines in the templates.

\subsection{Cross-correlation}
\label{sec:Cross-correlation}

\begin{figure*}
    \includegraphics[width=2\columnwidth]{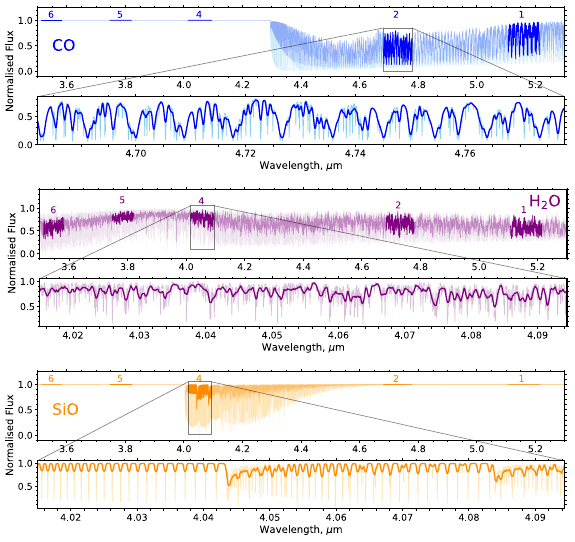}
    \caption{The normalised and continuum subtracted molecular models of CO (top), H$_2$O (middle), and SiO (bottom) from \textit{petitRADTRANS} used in this work. The models at the instrument resolution of R~=~100,000 are show in light shades, while the models broadened to the measured 22 km~s$^{-1}$ spin-rotation of $\beta$ Pic b (plotted in bold) demonstrate the reduction in both the number of lines and line depth due to rotational broadening. Highlighted regions of the spectra denote wavelengths covered by the five usable CRIRES+ M4368 detector orders. Insets: the detailed band structure of the order with the greatest contribution to the signal seen in the cross-correlation for each molecule.}
    \label{fig:models}
\end{figure*}

Following the removal of the stellar signal and telluric contamination, the residual spectra at each slit position are primarily composed of noise. The exception is the spatial position containing the planetary companion, which includes an additional contribution from the planetary spectrum, buried in the noise. To extract this weak signal (S/N~<~1 per line) we cross-correlate with model atmospheric spectra (see Section \ref{sec:Models}) through calculation of the Pearson correlation coefficient, $\rho$:
\begin{equation}
\rho_{X,Y} = \frac{C_{X,Y}}{\sqrt{C_{X,X} \cdot C_{Y,Y}}}
\end{equation}

where $\rho_{X,Y}$ is the Pearson correlation coefficient between two matrices $X$ and $Y$, and $C$ is the covariance. A cross-correlation function (CCF) is produced for each model (see Section \ref{sec:Results}) by calculating the correlation coefficient for each spatial position, across velocity shifts of $\pm$~210 km~s$^{-1}$, sampled at the CRIRES+ resolution element width of 1.5~km~s$^{-1}$. The planetary signal itself is Doppler shifted by a velocity shift $v_{\text{shift}}$:
\begin{equation}
v_{\text{shift}} = v_{\text{p}} + v_* + v_{\text{bary}}
\end{equation}

where $v_{\text{p}}$ is the planetary orbital velocity along the line of sight, $v_*$ the systemic velocity of the host star $\beta$ Pic A, and $v_{\text{bary}}$ the motion of the Earth relative to the solar barycentre at the time of observation. The planetary signal is therefore located at a unique spatial separation and velocity shift, which allows it to be distinguished from signals from $\beta$ Pic A and telluric contamination. In the stellar rest frame, residuals from $\beta$ Pic A are found at velocity shifts of 0~km~s$^{-1}$ at the stellar spatial position; and telluric contamination is found at velocity shifts of $\approx$ -13~km~s$^{-1}$ at all spatial separations. Additionally, the planetary velocity shift helps to delineate the planetary signal from any contaminating features from the inner regions of the $\beta$ Pic circumstellar debris disc. The total cross-correlation arrays are produced by summing the cross-correlation arrays from each frame and nod, following the application of a binary weighting to the orders based upon whether the model template contains spectral features at the corresponding wavelengths. Detector orders which are predicted to host model spectral lines are therefore given a weighting of one, while orders containing no molecular lines for a given model are given a weighting of zero, to prevent additional random noise contaminating the CCFs.

\begin{figure*}
    \includegraphics[width=2\columnwidth]{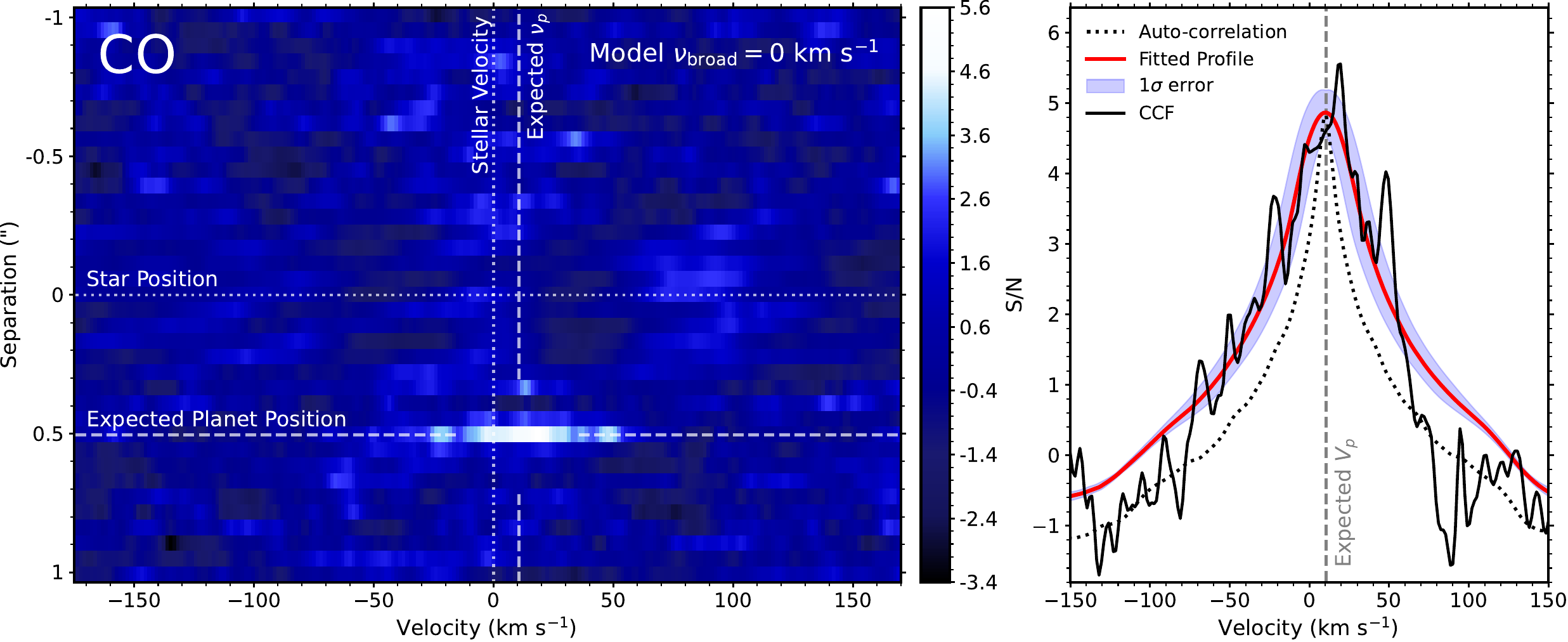}
    \caption{Left: The cross-correlation S/N of $\beta$ Pic b using a CO model at the instrument spectral resolution, as a function of spatial separation and velocity shift, centred at the systemic velocity of $\beta$ Pic A. We detect a clear CO signal from $\beta$ Pic b, at the expected planetary position (0.506\arcsec) and a velocity shift of 10$\pm$2~km~s$^{-1}$, using the first and second detector orders, centred at 4.73~$\mu$m and 5.16~$\mu$m. Right: The cross-correlation function at the planet spatial separation. The cross-correlation peak has a S/N~=~5.6, and is rotationally broadened with respect to the auto-correlation function of the model (indicating the signal expected from a non-rotating planet) by 21$\pm$5~km~s$^{-1}$. The fitted broadened profile is the result of cross-correlating the CO model template with a CO model template which has been rotationally broadened by 21~km~s$^{-1}$.}
    \label{fig:CO_CCF}
\end{figure*}

The calculation of a robust estimate of the S/N of a molecular detection from a CCF map requires specific care, as the detection significance is susceptible to statistically incorrect optimisation from the fine tuning of parameters \citep{Cabot2019}. We calculate the noise at each spatial position as the standard deviation of the CCF in that row, with the exception of the row containing the planet signal. Using our knowledge of the planet spatial position, we exclude a region of $\pm$70~km~s$^{-1}$ around the planet signal before calculating the row standard deviation, to avoid the peak biasing the estimate of the standard deviation of the row. The S/N is subsequently calculated in a row-wise fashion. We find this method to be more robust against fine tuning, although it potentially provides an underestimate of the S/N in the case of strong model auto-correlation functions, as seen for SiO (see Section \ref{sec:Results}).

\subsection{The impact of rotational broadening}
\label{sec:impact of rot broad}
$\beta$ Pic b is a freely rotating planet, with previously measured spin-rotations of 19.9$\pm$1.0~km~s$^{-1}$ \citep{Landman2023b} and 25$\pm$3~km~s$^{-1}$ \citep{Snellen2014}, which results in a broadening of its observed spectral lines (see Fig.~\ref{fig:models}). In order to measure this rotation, while also optimising our search for atmospheric species, we construct cross-correlation maps using two sets of models. First, we use models at the instrument resolution of R~=~100,000 to cross-correlate with the broadened planetary lines in the data. This provides a more conservative S/N detection, as the model templates have a higher effective spectral resolution than the data, but are more sensitive to the individual line profiles, and thus permits measurement of the spin-rotation and radial velocity of the planet (Sections \ref{sec:Results}, \ref{sec:Dynamics}, and \ref{sec:Orbital Solution}). Furthermore, the use of models at the instrument resolution reveals the underlying noise properties of the CCF in greater detail, allowing us to better distinguish between astrophysical and spurious signals. Using the measurements of rotational broadening based on the cross-correlation with the instrument resolution model templates, we subsequently broaden the molecular model templates by the measured rotational broadening for each molecule, and again cross-correlate with the data. The results for these detections are expected to show a higher S/N, as they provide a closer match to the broadened planetary lines in the data (\citealt{Spring2022}; see Section \ref{sec:Results}). However, when cross-correlating with a broadened model template neighbouring points in the CCF are highly correlated due to the width of spectral features in the template, and estimating the noise as the standard deviation of the CCF when sampled at the CRIRES+ resolution element width of 1.5~km~s$^{-1}$ is no longer appropriate. To resolve this issue we sample the CCF at every N\textsuperscript{th} point, with N tuned to $\sqrt{2}$~$v_{\text{rot}}$, where $v_{\text{rot}}$ is the measured velocity, and therefore the half width at half maximum (HWHM), of the planetary signal in the CCF \citep{Sebastian2024}. This represents the region over which we expect signals to be correlated, following the convolution of the model template and the planetary lines in the data, both of which are broadened by $v_{\text{rot}}$. The noise estimate of each row for the CCF at the instrument resolution element width is then calculated to be the standard deviation of the corresponding row in the down-sampled CCF.

\subsection{Measuring the rotational broadening}
\label{sec:measure of rot broad}

We measure the rotational broadening of $\beta$ Pic b for each molecular detection by fitting to the CCFs produced when using the model templates at the instrument resolution. First, we generate mock 1D CCFs by cross-correlating rotationally broadened model templates, representing the broadened planetary signal, with the unbroadened model used in the cross-correlation. To perform the broadening a rotation kernel with a limb-darkening coefficient of 0.3 is selected, representing mild planetary limb darkening \citep{Gray2008}. Limb-darkening is a strongly wavelength-dependent effect, which decreases in prominence towards longer wavelengths. The value used at \textit{M}-band wavelengths, calculated through comparison to brown dwarf analogues of $\beta$ Pic b \citep{Southworth2015}, is thus lower than previously retrieved limb-darkening coefficients for $\beta$ Pic b in the \textit{K}-band \citep{Landman2023b}. The resulting mock 1D CCFs are subsequently fit to the observed CCF peak using \textit{emcee} \citep{Foreman-Mackey2013}, with the posterior distributions providing estimates and errors for the radial velocity and planetary rotation velocity $v$sin(i). The broadened mock 1D CCFs are fit only to the central peak between $\pm$100~km~s$^{-1}$ to prevent biases introduced by fitting the structured auto-correlation wings to the noisy baseline. We note that, due to the broadening of the planetary signal, adjacent points in the velocity axis of the CCF are correlated. The use of $\chi^2$ in the MCMC fit in this case, therefore, may result in fitting errors which are underestimated in the reported uncertainties \citep{Sebastian2024}. 

\begin{figure*}
    \includegraphics[width=2\columnwidth]{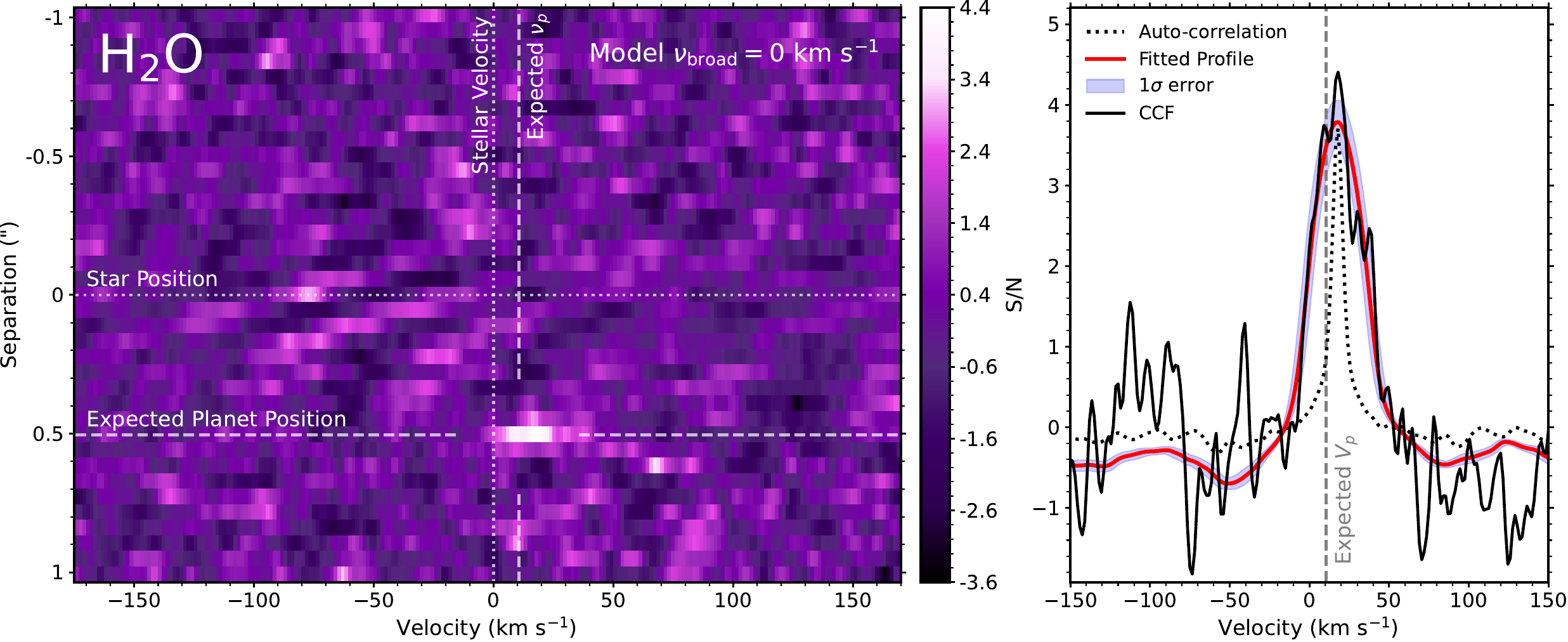}
    \caption{Left: The H$_2$O cross-correlation S/N of $\beta$ Pic b using models with the instrument spectral resolution, as a function of spatial separation and velocity shift, centred at the systemic velocity of $\beta$ Pic A. We see a weak H$_2$O signal in the expected region of the parameter space, when combining CCFs from the second, fourth, fifth, and sixth detector orders. Right: The cross-correlation function at the planet spatial separation. The signal has a maximum S/N = 4.4, short of the S/N~=~5 detection threshold, and a rotational broadening of 22$\pm$2 km~s$^{-1}$.
    }
    \label{fig:H2O_CCF}
\end{figure*}

\begin{figure*}
    \includegraphics[width=2\columnwidth]{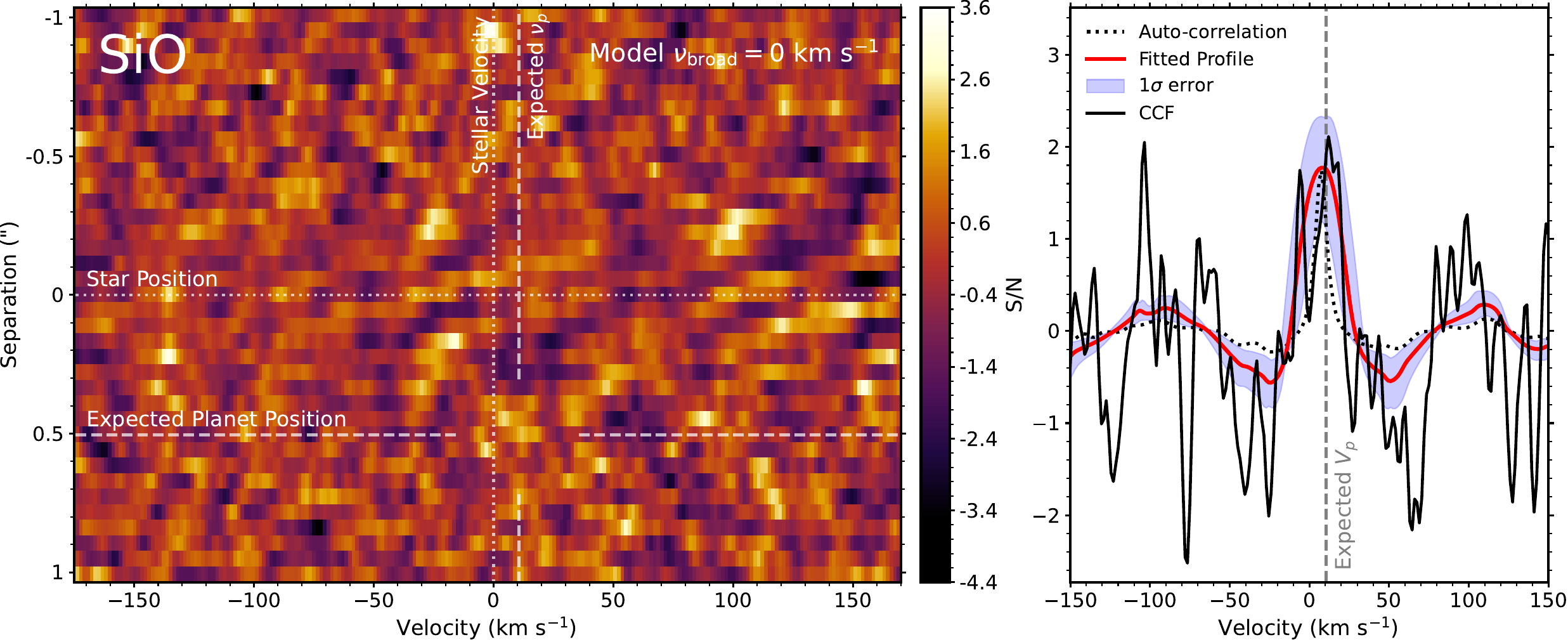}
    \caption{Left: The cross-correlation S/N produced using an SiO model at the instrument resolution, as a function of spatial separation and velocity shift, centred at the systemic velocity of $\beta$ Pic A. The CCF is produced using the fourth detector order and targets the 4~$\mu$m SiO band but shows no significant signal at the expected planet position. Right: The cross-correlation function at the planet spatial separation. There is a peak at the planet velocity shift with S/N=2.1 (below the S/N~=~5 threshold for robust detection). The peak does however match a broadening of 19$\pm$4 km~s$^{-1}$, consistent within the uncertainties of that measured from the CO and H$_2$O signals. Isolated signals with higher S/N are also present in the CCF away from the planet spatial and velocity separation.}
    \label{fig:SiO_CCF}
\end{figure*}

\section{Results}
\label{sec:Results}

We detect CO via cross-correlation with a model template at the instrument resolution, at the expected planetary position (0.506\arcsec), with a S/N~=~5.6, and localised at 10$\pm$2 km~s$^{-1}$, consistent with the expected planetary radial velocity (Fig.~\ref{fig:CO_CCF}). Cross-correlating with the broadened model template gives an enhanced S/N~=~6.6, strengthening the detection. Our weighting criteria (see Section \ref{sec:Cross-correlation}) includes signals from the first and second detector orders, with the second order at 4.73~$\mu$m containing the majority of the signal. The CCF contains residual noise structures, most notably banding in the main stellar trace region, and isolated noisy regions reaching S/N $>$~3 away from the planet location. The localisation, width, and S/N of the peak in the CCF, however, provide a strong detection, and a clear validation of the functionality of HRCCS techniques in the \textit{M}-band. Contamination from residual telluric lines are expected to manifest in the CCF as signals at a velocity shift of $\approx$ -13~km~s$^{-1}$ in the stellar rest frame, and to be present across all spatial separations. These signals are seen to dominate the CCF when a small number of PCA components are removed but are removed with increasing PCA components. The CCF also shows `stripes' in velocity space, most prominently at the spatial position of the star, which result from persistent residuals from imperfect stellar subtraction, driven by instrument systematics (see Section \ref{sec:Post-Processing}). While these stripes are also present in the CCF produced using a H$_2$O model template they do not impact the signal at the planet spatial position, which is thermal noise dominated. 

The signal is visibly broadened with respect to the auto-correlation of the model template, the expected cross-correlation result for a non-rotating planet. The signal produced from cross-correlation with the instrument resolution CO model template is best fit by a rotational broadening of 21$\pm$5~km~s$^{-1}$, consistent with both the 25$\pm$3~km~s$^{-1}$ \citet{Snellen2014} in the \textit{K}-band using the original CRIRES, and the recent 19.9$\pm$1.0~km~s$^{-1}$ CRIRES+ measurement of \citet{Landman2023b} in the \textit{K}-band. The detection of a signal at 4.73~$\mu$m is challenging, as the second detector order displays strong telluric contamination from H$_2$O, O$_3$, and a minor contribution from CO, alongside high thermal background noise. This detector order, however, also presents a number of key observational advantages. Firstly, $\beta$ Pic b has a star-planet blackbody contrast ratio of 9.4$\times$10$^{-4}$ at 4.73~$\mu$m, a 40$\%$ increase on the contrast ratio at 3.5~$\mu$m. Furthermore, the MACAO achieves a dramatic suppression of the starlight at the planet position, and an AO suppression factor of $\approx$100 is measured at 4.73~$\mu$m. As the molecular species with the strongest opacity at this wavelength, paired with a high predicted equilibrium abundance, and a spectrum displaying deep, well-spaced ro-vibrational absorption lines, CO is observable in $\beta$ Pic b at 4.73~$\mu$m. Finally, it is worth highlighting that the forest of telluric lines present across the second order, while challenging to remove in the data post-processing, also enable a robust wavelength solution, vital for the success of HRCCS.

While H$_2$O is predicted to be readily observable, with a high equilibrium abundance and spectral features spanning all usable detector orders, we see a comparatively weak signal in our CCF at S/N = 4.4 when using models at the instrument resolution, short of the S/N~=~5 detection threshold (Fig.~\ref{fig:H2O_CCF}). Additionally, there are isolated regions approaching S/N levels comparative to the planet signal. Cross-correlating with the broadened model template, however, produces a firm detection of S/N~=~5.7. The CCFs produced using H$_2$O model templates are constructed using the second, fourth, fifth, and sixth detector orders; the first detector order is excluded due to the exceptionally high contamination from telluric H$_2$O at 5.2 $\mu$m. The CCF produced through cross-correlation with a H$_2$O model at the instrument resolution shows a rotational broadening of 22$\pm$2~km~s$^{-1}$, consistent with the CO broadening measurement. The lower S/N detection H$_2$O when cross-correlating with models at the instrument resolution is likely a result of the molecular band structure of H$_2$O. The H$_2$O model spectrum contains a high density of shallow spectral lines, which are readily buried in the thermal background noise and eroded by the PCA-based removal of systematic trends. The cause of the increase in S/N when using the broadened model is twofold; first, the template spectra more closely match the data, in which the planet spectral lines are broadened, and second, the spurious noise features in the CCF are smeared out, reducing the impact of isolated noise sources.

\begin{figure*}
    \centering
    \includegraphics[width=2\columnwidth]{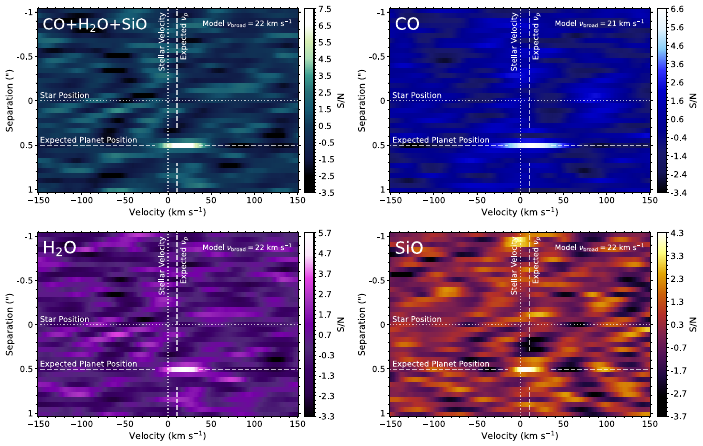}
    \caption{Cross-correlation maps produced using models broadened to the measured rotational broadening of each molecule. These CCFs are predicted to show stronger detections as the models used more closely match the broadened planetary lines in the data, and the influence of noise sources on the scale of a single resolution element in the CCF are suppressed. Most notably, when cross-correlating with an SiO model broadened to the 22~km~s$^{-1}$ rotational broadening measured from the H$_2$O CCF, the evidence for the tentative SiO signal increases to an S/N~=~4.3, while other regions are suppressed, suggesting a potential planetary origin. The CO signal is boosted to S/N~=~6.6 and the H$_2$O signal is now above the S/N~=~5 threshold for a detection, with an S/N~=~5.7. Cross-correlation with a broadened model containing CO, H$_2$O, and SiO produces the strongest detection of all models tested, with a S/N~=~7.5. The CCFs produced using the H$_2$O model and the model containing all three molecular opacities show an offset from the expected velocity, which is discussed further in Section \ref{sec:Orbital Solution}. To reduce the impact of correlated signals the noise estimates for these CCFs are measured from a down-sampled CCF, with the planet signal excluded (see Section \ref{sec:impact of rot broad}). For clarity, and following the convention for HCS, the CCFs are presented here at the 1.5~km~s$^{-1}$ CRIRES+ pixel resolution.}
    \label{fig:All_broad}
\end{figure*}

\begin{table}
	\centering
	\caption{Summary of the measurements made in the work. S/N$_{\text{inst}}$ and S/N$_{\text{broad}}$ denote the peak S/N achieved when cross-correlating with the molecular model templates at the CRIRES+ resolution, and with those that have been broadened to the measured planetary rotational velocity ($v_{\text{rot}}$), respectively. The uncertainties for $v_{\text{rot}}$ and $v_{\text{p}}$ (the planetary radial velocity; predicted to be 10.6$\pm$0.1 km~s$^{-1}$, \citealt{Wang2021b}) are estimated using an MCMC fit to the CCF (see Fig.~\ref{fig:CO_Corner}). Note that $\lambda$ range represents the non-continuous wavelength coverage used to detect the molecule, while orders denote which echelle orders from CRIRES+ were used in the cross-correlation. \label{results_table}}
	\label{tab:results_table}
	\begin{tabular}{ccccccc} 
		\hline
            Model & $\lambda$ & Orders &S/N$_{\text{inst}}$ & S/N$_{\text{broad}}$ & $v_{\text{rot}}$ & $v_{\text{p}}$\\
            &$\mu$m&&&&km~s$^{-1}$&km~s$^{-1}$\\
            
		\hline
            \hline
            CO & 4.7--5.2 & 1,2 & 5.6 & 6.6 & 21$\pm$5& 10$\pm$2 \\
            H$_2$O & 3.5--4.8 & 2,4,5,6 & 4.4 & 5.7 & 22$\pm$2&18$\pm$2$^a$\\
            SiO & 4.0--4.1 & 4 & 2.1 & 4.3 & 19$\pm$4$^b$& 7$\pm$2\\
            All & 3.5--4.8 & 2,4,5,6 & 5.2 & 7.5 & - & -\\
		\hline
  \hline
	\end{tabular}
\begin{flushleft}
\footnotesize{$^a$ The measurement of $v_{\text{p}}$ for the H$_2$O detection shows a large offset from the predicted planetary velocity, which is discussed further in Section \ref{sec:Orbital Solution}.\\}
\footnotesize{$^b$ The value of $v_{\text{rot}}$ for the SiO detection is measured from a low S/N$_{\text{inst}}$ signal, and is therefore highly uncertain.}
\end{flushleft}
\end{table}

The cross-correlation using an SiO model template at the instrument resolution produces a noisy CCF with no prominent detection peak (Fig.~\ref{fig:SiO_CCF}). The signal that is aligned with the planet spatial separation and velocity shows a broadening of 19$\pm$4~km~s$^{-1}$ compared to the model auto-correlation function, and a peak S/N~=~2.1. The global properties of the CCF would indicate that this observed signal is consistent with spurious noise as there are isolated signals with higher S/N present in the CCF which do not fall at the planetary position. Additionally, the negative values in the CCF reach S/N~=~-3.7 and while these strong anti-correlation values could be attributed to the intrinsic auto-correlation of the SiO model, the Gaussian statistics of the CCF imply that false positive signals of the same significance, but in positive S/N, are equally likely to occur \citep{Cabot2019}. If we anticipate false positives of S/N~=~+3.7, then we must also consider that the 2.1 S/N `planet' signal falls below this threshold.

\begin{figure}
    \includegraphics[width=\columnwidth]{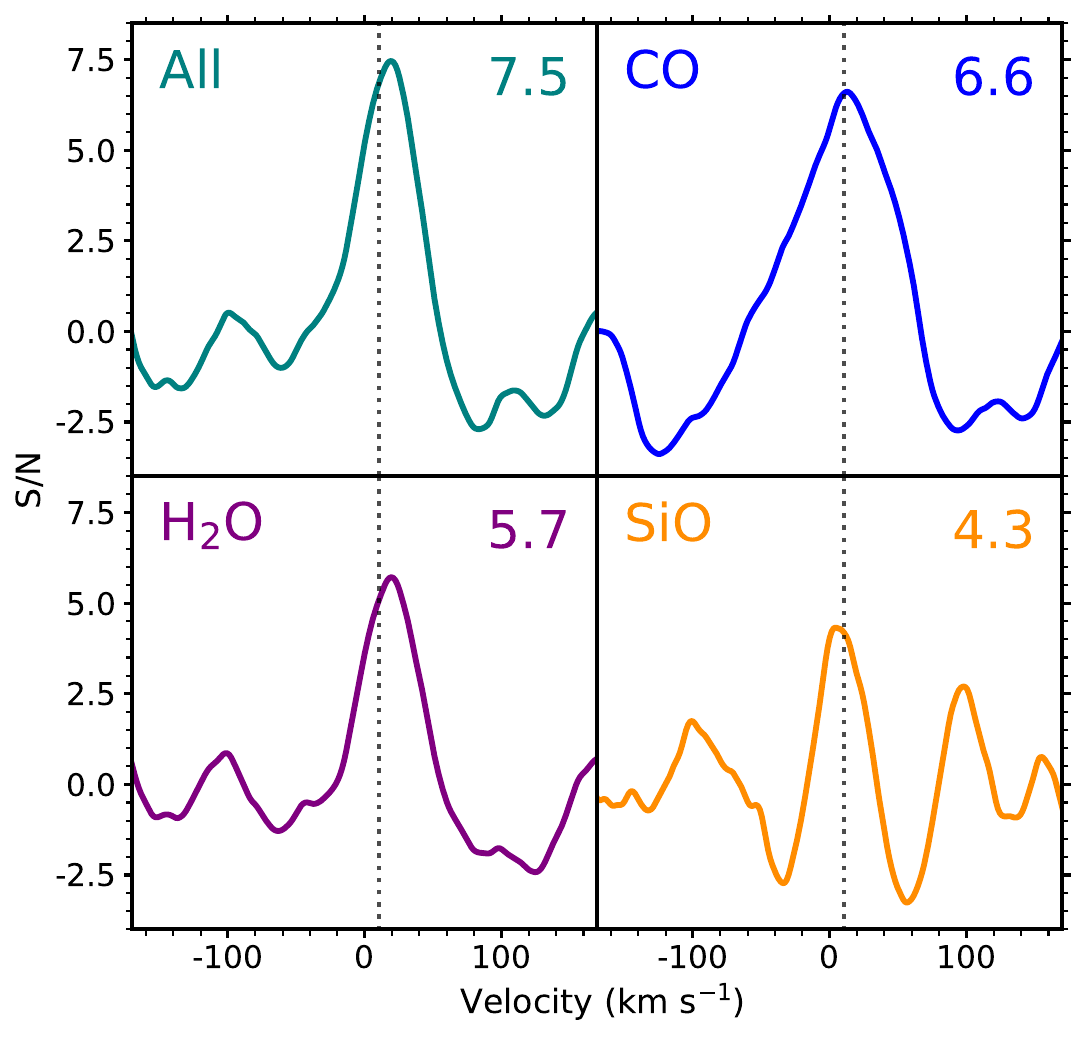}
    \caption{The cross-correlation functions of the broadened model CCFs at the planet spatial separation, in the rest frame of $\beta$ Pic A. The top right number denotes the peak S/N of each signal. The model `all', contains opacities from CO, H$_2$O, and SiO. The dotted line marks the expected planetary orbital velocity.}
    \label{fig:all_broad_CCF_slices}
\end{figure}

However, cross-correlating with an SiO model broadened to the 22~km~s$^{-1}$ rotational broadening measured from the H$_2$O CCF shows a significant enhancement in the S/N compared to the use of models at the instrument resolution, and the SiO CCF is strongly peaked at the planet position, providing marginal evidence for SiO at an S/N~=~4.3 (Fig.~\ref{fig:All_broad}). This boost in S/N when using a broadened model, which better matches the predicted planet properties, supports a planetary origin for the SiO signal. This interpretation is strengthened when paired with a rotational broadening measurement of 19$\pm$4~km~s$^{-1}$, consistent with the CO and H$_2$O detections. In addition, the CCF produced through cross-correlation with the broadened SiO template spectra displays subsidiary structure at velocity shifts of $\pm$100~km~s$^{-1}$ (see Fig.~\ref{fig:all_broad_CCF_slices}). We explore the possibility that these signals are aliases of the model template, and indeed the model auto-correlation function is structured and shows significant peaks at velocity shifts of $\approx$ $\pm$110~km~s$^{-1}$, as a result of the regular spacing of molecular lines in the SiO band structure (see Fig.~\ref{fig:models}). This potential model aliasing lends greater confidence to a planetary origin of the SiO signal, but is difficult to directly incorporate into the S/N estimate \citep{Esparza-Borges2023}. Furthermore, the isolated peaks seen in the CCF produced when cross-correlating with the instrument resolution SiO model are suppressed when using the broadened model, while the planet signal is enhanced. Following the S/N~=~5 detection criteria, however, the S/N ratio of the SiO peak remains too low to claim a robust detection but warrants further data. For completeness, in Section \ref{sec:Discussion} we will explore the implications of a SiO signal from $\beta$ Pic b, with the caveat that these results provide only marginal evidence of SiO, and motivate future work to investigate this further.

When cross-correlating with the model containing CO, H$_2$O, and SiO, broadened to the planetary rotation velocity, a signal at S/N~=~7.5 is produced (Fig.~\ref{fig:All_broad}). As this model contains all three opacities, and is broadened to the measured planetary rotation velocity, it represents our best estimate of the planet's \textit{L} and \textit{M}-band spectrum. A summary of the detected molecular species is presented in Table~\ref{tab:results_table}. The remaining molecular templates tested (PH$_3$, H$_2$S, CH$_4$, NH$_3$) show no signals at the planetary location when cross-correlated with the processed data, in line with the sensitivity of our data and the predicted equilibrium abundances (see Appendix~\ref{sec:App}, Fig.~\ref{fig:all_other_ccfs}).

\section{Discussion}
\label{sec:Discussion}
This work tests the feasibility of HRCCS techniques in a new spectral regime, the thermal noise dominated \textit{M}-band between 3.51--5.21~$\mu$m. We find it is possible to detect molecular species, most prominently CO, although the residual noise in the CCFs when using models at the R~=~100,000 instrument resolution appears higher than is experienced at shorter wavelengths.  Interestingly, when using broadened models, we achieve a similar detection significance and broadening properties for CO as found in one hour of \textit{K}-band CRIRES observations by \citet{Snellen2014}, but in a significantly more challenging wavelength regime, and requiring approximately twice the observing time. This demonstrates the increase in data quality for HRCCS from the CRIRES upgrade, most notably from the improved wavelength coverage, but is also testament to the advances in post-processing techniques for high-resolution data over the past decade. Here, we will first discuss the implications of these results for the chemistry, dynamics, and orbital solution of $\beta$ Pic b. Additionally, we discuss the noise properties of HRCCS in the \textit{M}-band regime, observational prospects in the \textit{M}-band using existing instrumentation, and finally the future of HRCCS in the \textit{M}-band with the ELTs.

\subsection{The atmosphere and clouds of \texorpdfstring{$\beta$}{Beta} Pic b}
\label{sec:Atmosphere of Beta Pic b}
 
High-resolution studies of the atmospheric composition of $\beta$ Pic b have provided direct detections of CO and H$_2$O \citep{Snellen2014, Hoeijmakers2018, Landman2023b}, while photometric direct imaging studies, which have sampled the spectral energy distribution of $\beta$ Pic b from 1--5~$\mu$m, favour a low surface gravity (log(g)~$\approx$~3.5--4.5) and effective temperatures in the range of 1650 to 1800~K \citep{Bonnefoy2013,Currie2013,Morzinski2015,Stolker2020}. Our \textit{L} and \textit{M}-band detections of CO and H$_2$O conform to and corroborate this observational consensus but do not provide additional constraints on the abundances of CO and H$_2$O in $\beta$ Pic b as the S/N used in this work is largely invariant to abundance changes. The high effective temperature, T$_{\text{eff}}$~$\approx$~1700~K, and strong internal residual heat from formation in $\beta$ Pic b ensure that, in equilibrium, the atmospheric carbon budget is stored almost exclusively in CO \citep{Lodders2002}. H$_2$O appears to be ubiquitous in giant planet atmospheres and is predicted to remain a dominant carrier of oxygen across all pressure levels in $\beta$ Pic b, with VMR close to the CO abundance.

Silicon monoxide (SiO) is predicted to be the most abundant silicon bearing gas in the atmospheres of hot (T$_{\text{eff}}$ $\gtrsim$ 900~K) young giant planets \citep{Sharp2007}. In higher pressure regions (log($P$) $\gtrsim$ 2) silane (SiH$_4$) dominates, while at lower temperatures SiH$_2$F$_2$ or SiH$_3$F become the dominant carriers of silicon \citep{Visscher2010}. The gas chemistry and abundance of SiO in hot giant planets is influenced primarily by magnesium-silicate cloud condensates, and SiO will most readily condense to form Mg$_2$SiO$_4$ (forsterite) and MgSiO$_3$ (enstatite), providing an atmospheric sink of gaseous SiO \citep{Gao2021}. The abundance of gaseous SiO above a magnesium-silicate cloud deck is quenched and tied to the efficiency with which condensation removes silicon from the gas phase, while below a cloud deck, or in the absence of magnesium-silicate clouds, SiO is predicted to be the most abundant Si-bearing gas in the atmosphere, with an abundance of log(X$_{\text{SiO}}$)~$\approx$~-4.20~+~[Fe/H] \citep{Visscher2010}. On $\beta$ Pic b, therefore, the chemistry of SiO is explicitly linked to cloud formation. Thick clouds have broadly been favoured by photometric and spectroscopic studies of $\beta$ Pic b \citep{Morzinski2015,Chilcote2017,GRAVITY2020, Stolker2020}, although measurements of cloud properties suffer from a strong degeneracy with metallicity \citep{Landman2023b}. Two key classes of clouds have been considered to contribute to the observed cloudy properties of $\beta$ Pic b; magnesium-silicate clouds, and Fe hazes \citep{GRAVITY2020}. In the presence of a magnesium-silicate cloud deck, SiO is sequestered from the gas phase into clouds and the observed abundance of gaseous SiO above the cloud deck is severely depleted. However, the marginal evidence for SiO in the \textit{M}-band data requires a suitably high abundance to be observable, and therefore this signal is in tension with previously predicted magnesium-silicate cloud decks. We consider three degenerate scenarios which could reconcile the marginal \textit{M}-band SiO signal with our present understanding of cloud properties on $\beta$ Pic b.

First, the signal of SiO may imply the absence of magnesium-silicate condensation in the atmosphere at all pressure levels. In the absence of magnesium-silicate clouds, then Fe hazes are likely the key contributor to the scattering properties of $\beta$ Pic b. Alternatively, the marginal SiO detection allows for magnesium-silicate clouds with an incomplete or non-uniform spatial distribution across the observed planetary hemisphere. In this `patchy cloud' scenario the gaseous SiO would lie in atmospheric layers below the magnesium-silicate cloud deck, but gaps in the clouds allow emission from these layers to radiate outwards. Patchy clouds have been proposed as a cause of the high photometric variability of planetary-mass objects near the L-T transition (e.g. \citealt{Vos2023}), but $\beta$ Pic b more closely resembles earlier L dwarfs where the cloud coverage is expected to be more homogeneous (e.g. \citealt{Brock2021}). The contribution function for SiO at 4~$\mu$m reaches a maximum photospheric pressure of 1~bar, deeper in the atmosphere than the predicted cloud base, and therefore these observations would be sensitive to atmospheric layers below the hypothetical patchy magnesium-silicate cloud deck proposed in this scenario (see Fig.\ref{fig:fastchem}). The photometric variability resulting from patchy cloud structures over the rotation period of the planet opens the prospect of testing this scenario with time variability studies (e.g. \citealt{Zhou2016,Zhou2020,Wang2022,Sutlieff2023}). Finally, the cloud particle size distribution may result in a reduced 4~$\mu$m opacity, and thus transparency at \textit{L} and \textit{M}-band wavelengths that allows us to see just below the cloud base where SiO should be at its most abundant \citep{Visscher2010}. Both an optically thick distribution of sub-micron scale grains and an optically thin layer of particles with larger radii could result in the required 4~$\mu$m transparency.

The existence and exact composition of clouds in giant planets enables us to understand both the temperature switch at which clouds of different species condense, and the formation sites of migrated giant planets. For the latter, previous works have targeted C/O ratios from gaseous volatiles (e.g. CO, H$_2$O delivered through ices and gas in the protoplanetary disc) to trace formation sites \citep{Oberg2011,Oberg2023}, including for $\beta$ Pic b \citep{GRAVITY2020,Landman2023b}. This approach can lead to degeneracies due to varying thermal and chemical processing at the origin of the gaseous volatiles \citep{Mordasini2016, Lichtenberg2021}. Refractory material (e.g. Si and Fe) however, can break the degeneracy as it remains in the solid phase throughout most of the protoplanetary disc, and thus their atmospheric abundance ratios in exoplanetary atmospheres more reliably reflect the solid-to-gas accretion ratios during formation \citep{Chachan2023}. The hint of SiO in the \textit{M}-band therefore offers the possibility of using new abundance ratios of refractory elements, e.g. the Si/H, O/Si, and C/Si ratios, to place constraints on the ratio of ices to rocks accreted during the planet's formation and where, when, and how in the protoplanetary disc this occurred. To date, gaseous SiO has not been uniquely detected in an exoplanet atmosphere, although HST/WFC3 UVIS transit spectroscopy of the ultra-hot Jupiter WASP-178b suggested the presence of either SiO or a super-solar abundance of Mg I and Fe II \citep{Lothringer2022}, with strong theoretical motivation for both SiO and Mg to be present as cloud precursors. Spatially resolved high-resolution spectroscopy in the \textit{M}-band, targeting the strong 4~$\mu$m SiO bands (Fig.~\ref{fig:models}), therefore offers a new window on silicate chemistry and cloud formation in sub-stellar objects and giant exoplanets.

\subsubsection{Non-detections} 
We searched for PH$_3$ (phosphine), which has a very small predicted abundance in equilibrium, to investigate disequilibrium chemistry. In disequilibrium PH$_3$ is expected to exhibit features in $\beta$ Pic b in the \textit{M}-band, due to e.g. dredge up from deeper layers, as seen in Jupiter’s atmosphere \citep{Fletcher2009, Moses2016}. We also searched for H$_2$S as a probe of photochemical disequilibrium processes \citep{NowakG2020,Zahnle2016}. We obtain no significant detection of either PH$_3$ or H$_2$S, but due to the sensitivity of our observations this does not place meaningful constraints on their presence in the atmosphere of $\beta$ Pic b. Additionally, we searched for signals from CH$_4$ and NH$_3$, and see no significant detections, consistent with the non-detections from \citet{Hoeijmakers2018} as these molecules are not expected to be present in significant quantities under the predicted carbon and nitrogen enrichment of the atmosphere \citep{Moses2016}. We see no signals in any of the CCFs from the planet $\beta$ Pic c, which would be expected at a velocity shift of -11$\pm$1~km~s$^{-1}$, and a separation of -0.105\arcsec. Signals at this separation are dominated by contamination from the star (see Section \ref{sec:M-band}).

\subsection{Rotation and dynamics}
\label{sec:Dynamics}

\begin{figure}
    \includegraphics[width=\columnwidth]{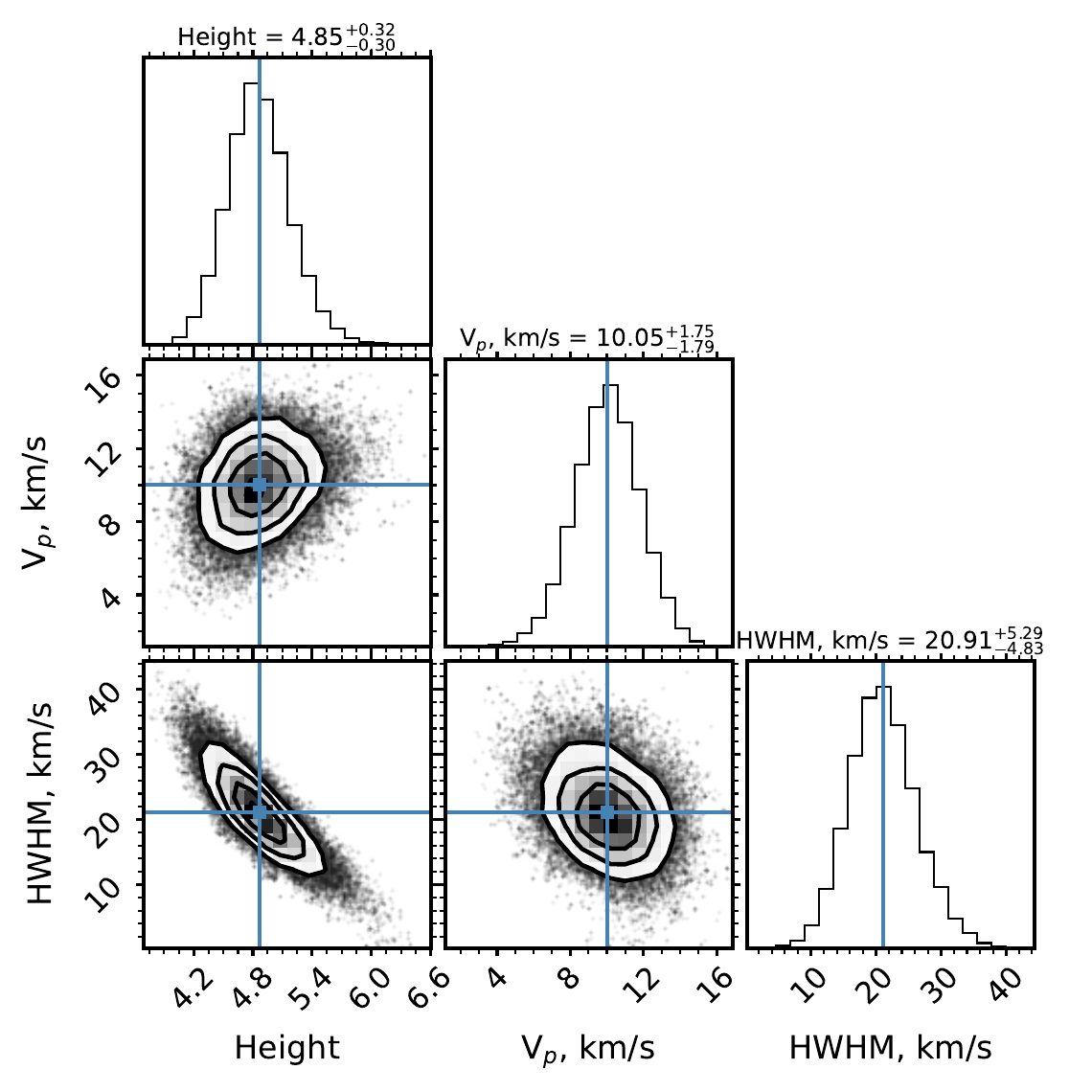}
    \caption{The Posterior distributions from the MCMC fit to the CCF produced using a CO model template at the instrument resolution. The height, position ($v_p$), and model broadening half width at half maximum (HWHM) parameters are fit. The fitted models are produced through cross-correlation of the CO model template at the instrument resolution with a rotationally broadened copy of itself. Note the degeneracy between the fitted peak height and fitted width (HWHM), which introduces considerable uncertainty in the measurement of the planet $v$sin($i$).}
    \label{fig:CO_Corner}
\end{figure}

The most precise \textit{M}-band measurement of the rotational broadening, from the H$_2$O signal, $v$sin($i$)~=~22$\pm$2~km~s$^{-1}$, is consistent within 1$\sigma$ error with the 25$\pm$3~km~s$^{-1}$ constraint of \citet{Snellen2014}, the CRIRES+ \textit{K}-band measurement of 19.9$\pm$1.0~km~s$^{-1}$ \citep{Landman2023b}, and the recent 25$^{+5}_{-6}$~km~s$^{-1}$ obtained through reanalysis of R~$\approx$~4000 SINFONI data \citep{Kiefer2024}. We report a rotation period for $\beta$ Pic b of 8$\pm$1 hours, when using planetary parameters from \citet{Landman2023b}\footnote{R$_p$~=~1.4$\pm$0.1~R$_J$, i~=~88.95$\pm$0.1$^{\circ}$, zero planetary obliquity.}. The precision of the spin-rotation measurement in this work, which is lower than for the comparative CRIRES+ \textit{K}-band measurement, is a direct consequence of the lower S/N detection using the model template at the instrument resolution, in which the measurement of the broadening is more susceptible to residual noise structures in the CCF. Additionally, our fitting suffers a degeneracy between peak height and fitted width, most prominent in the fitting of the CO CCF (see Fig.~\ref{fig:CO_Corner}).
Differences in fitting methods could also cause discrepancies in the measured rotational broadening and we therefore use the \textit{fastRotBroad} function from pyAstronomy\footnote{\url{https://github.com/sczesla/PyAstronomy}} \citep{pyastrnomy} to perform the rotational broadening, consistent with \citet{Landman2023b}. The fitting of rotational broadening in \citet{Landman2023b}, however, is wrapped within the full retrieval of atmospheric parameters, and is therefore not fully replicated here. 

\begin{figure}
    \includegraphics[width=1\columnwidth]{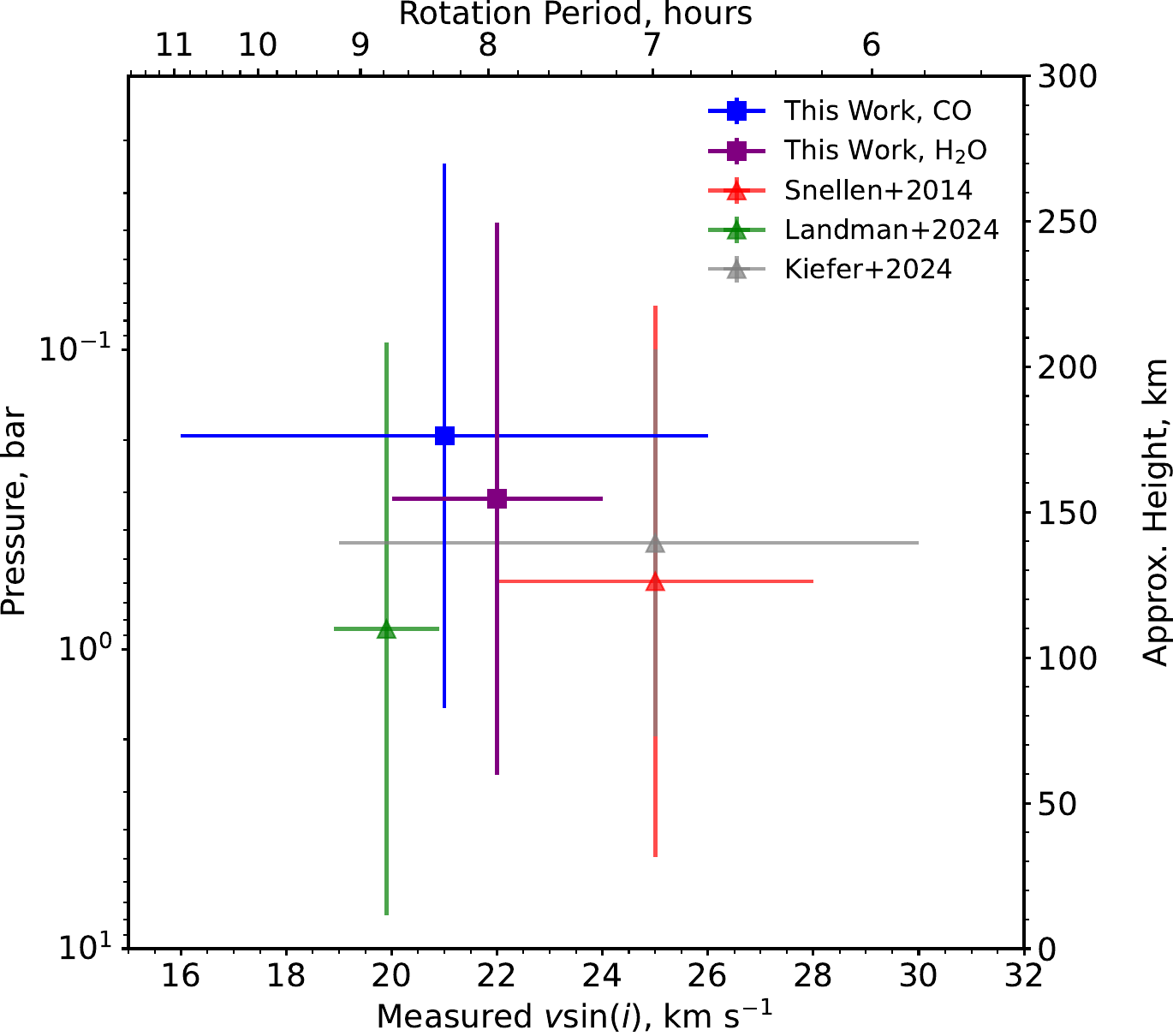}
    \caption{Comparison of the spin-rotation measured from the \textit{M}-band molecular detections in this work to previous studies of $\beta$ Pic b in the \textit{K}-band. The measurements of rotational broadening in this work are all consistent within 1$\sigma$ error and further observations at a higher S/N, alongside more detailed modelling of atmospheric contribution, would be required to probe vertical variations in rotation. Height is approximated isothermally from a base pressure level of 10 bar, and rotation period calculated consistently with \citet{Landman2023b}. The broadening measurement of SiO is excluded from this figure as it is measured from a low S/N signal.}
    \label{fig:windspeed}
\end{figure}

\citet{Snellen2014}, \citet{Landman2023b}, \citet{Kiefer2024}, and this study are sensitive to emission from different, but overlapping, atmospheric layers of $\beta$ Pic b. \citet{Landman2023b} measures the rotational broadening from a detection dominated by molecular H$_2$O found at 2.0$^{+7.0}_{-1.9}$ bar, while \citet{Snellen2014} (observing using the narrower wavelength coverage of the original CRIRES) primarily detect CO, which displays a \textit{K}-band contribution function that peaks at 0.50$^{+3.50}_{-0.43}$ bar (Fig.~\ref{fig:windspeed}). \citet{Kiefer2024} detect both CO and H$_2$O, probing similar atmospheric regions in the \textit{K}-band. In the \textit{M}-band we are sensitive to CO and H$_2$O emission slightly higher in the atmosphere than the previous studies in the \textit{K}-band (see Fig.~\ref{fig:fastchem}).

Comparing the measurements of spin-rotation of $\beta$ Pic b with the atmospheric layers probed by each study, we see each measurement is broadly consistent (Fig.~\ref{fig:windspeed}). If large deviations in measured velocity with altitude were detected it could suggest an extreme vertical wind shear in the atmosphere of $\beta$ Pic b, analogous to that seen on Neptune (2.2 m/s/km, \citealt{Tollefson2018}), albeit orders of magnitude stronger. At present, however, the uncertainties in our measurements of $v$sin(i) are too large to draw robust statistical conclusions. The largest uncertainties originate from estimating the model dependent atmospheric pressure level probed by each molecule, and from the error in determining the broadening of the CCF peaks. Together, these currently prevent detailed conclusions of the vertically resolved atmospheric velocity structure. Nonetheless, extending the wavelength coverage of HCS into the \textit{M}-band opens the opportunity to study the vertical rotation structure in widely separated giant planets and brown dwarf atmospheres by probing new atmospheric regions with HRCCS. Precise measurements of rotational broadening of different molecules across different spectral regions can be used to search for trends of velocity variation with height and could be leveraged to map the vertical wind structure in exoplanet atmospheres. The \textit{M}-band measurements of rotational broadening made in this work join the small number of spectroscopic measurements of the spin of non-tidally locked planetary-mass companions, including those made using HCS and CRIRES/VLT \citep{Snellen2014,Schwarz2016}, CRIRES+/VLT (\citealt{Landman2023b}; this work), NIRSPEC/Keck \citep{Bryan2018,Xuan2020}, SINFONI/VLT \citep{Kiefer2024}, and high-resolution fibre-fed spectroscopy using KPIC/Keck \citep{Wang2021b}.

\subsection{Orbital solution}
\label{sec:Orbital Solution}

Accounting for the 20.0$\pm$0.7~km~s$^{-1}$ systemic velocity of $\beta$ Pictoris A \citep{Gontcharov2006}, and the Earth's barycentric velocity, we detect the CO signal at a radial velocity of 10$\pm$2 km~s$^{-1}$, consistent with the predicted planetary orbital velocity of 10.6$\pm$0.1 km~s$^{-1}$ \citep{Lacour2021,Wang2021a}. The SiO signal peaks at a velocity shift of 7$\pm$2~km~s$^{-1}$, within 2$\sigma$ error of the theoretically predicted planetary velocity. The H$_2$O detection, however, has a measured velocity shift of 18$\pm$2~km~s$^{-1}$, where the $\pm$2~km~s$^{-1}$ uncertainty is the statistical error from the fitting to the CCF. This error does not, however, reflect any systematic uncertainties present in the data, which have the potential to bias the fitting and produce the 7~km~s$^{-1}$ red-shift from the predicted planetary velocity. The measurement of radial velocities from HCS CCFs is inherently challenging as it requires the determination of the peak of broadened signals in noisy data. For example, \citet{Landman2023b} measure the radial velocities of 40 individual exposures in the \textit{K}-band and, despite measuring individual radial velocities to precisions of $\pm$2~km~s$^{-1}$, observe an intrinsic scatter in the distribution of measured radial velocities for $\beta$ Pic b, with a 1$\sigma$ width of $\pm$2.3~km~s$^{-1}$ and maximum range of $\pm$5~km~s$^{-1}$. The CCFs in the \textit{M}-band suffer from more prominent systematic noise than those in the \textit{K}-band, and therefore it is likely that the measured offset for H$_2$O falls within the 2$\sigma$ bounds of the intrinsic scatter expected for the measurement of RVs by this method. However, we require the sum of all \textit{M}-band exposures to detect the planet signal and therefore cannot probe frame-by-frame contributions to the measured radial velocity value. Additional evidence to suggest an unphysical origin of the radial velocity offset is provided by the high S/N \textit{K}-band detection of H$_2$O in an overlapping pressure range to our \textit{M}-band detection, that shows no significant offset in radial velocity \citep{Landman2023b}. While the offset of the H$_2$O signal necessitates a critical evaluation of the confidence in the H$_2$O signal, a number of factors including the S/N~$>$~5, the spatial localisation of the H$_2$O signal, and the consistent measurement of rotational broadening all support a robust detection of H$_2$O using our \textit{M}-band data. Crucially, while the exact centre of the H$_2$O signal appears to be offset, the S/N is simply defined as how far the peak extends above the standard deviation of the CCF. We can therefore be confident in the significance of the H$_2$O signal.

Despite their large errors, these radial velocity measurements of $\beta$ Pic b, alongside those of \citet{Snellen2014} and \citet{Landman2023b}, may help constrain its eccentric orbit \citep{Nowak2020} and further constrain the dynamical mass of this directly imaged young planet \citep{Lacour2021} for bench-marking theoretical evolution models \citep{Baraffe2003}. However, this is beyond the scope of this work.

\subsection{Implications for future studies in the \textit{M}-band}
\label{sec:M-band}

\begin{figure}
    \includegraphics[width=\columnwidth]{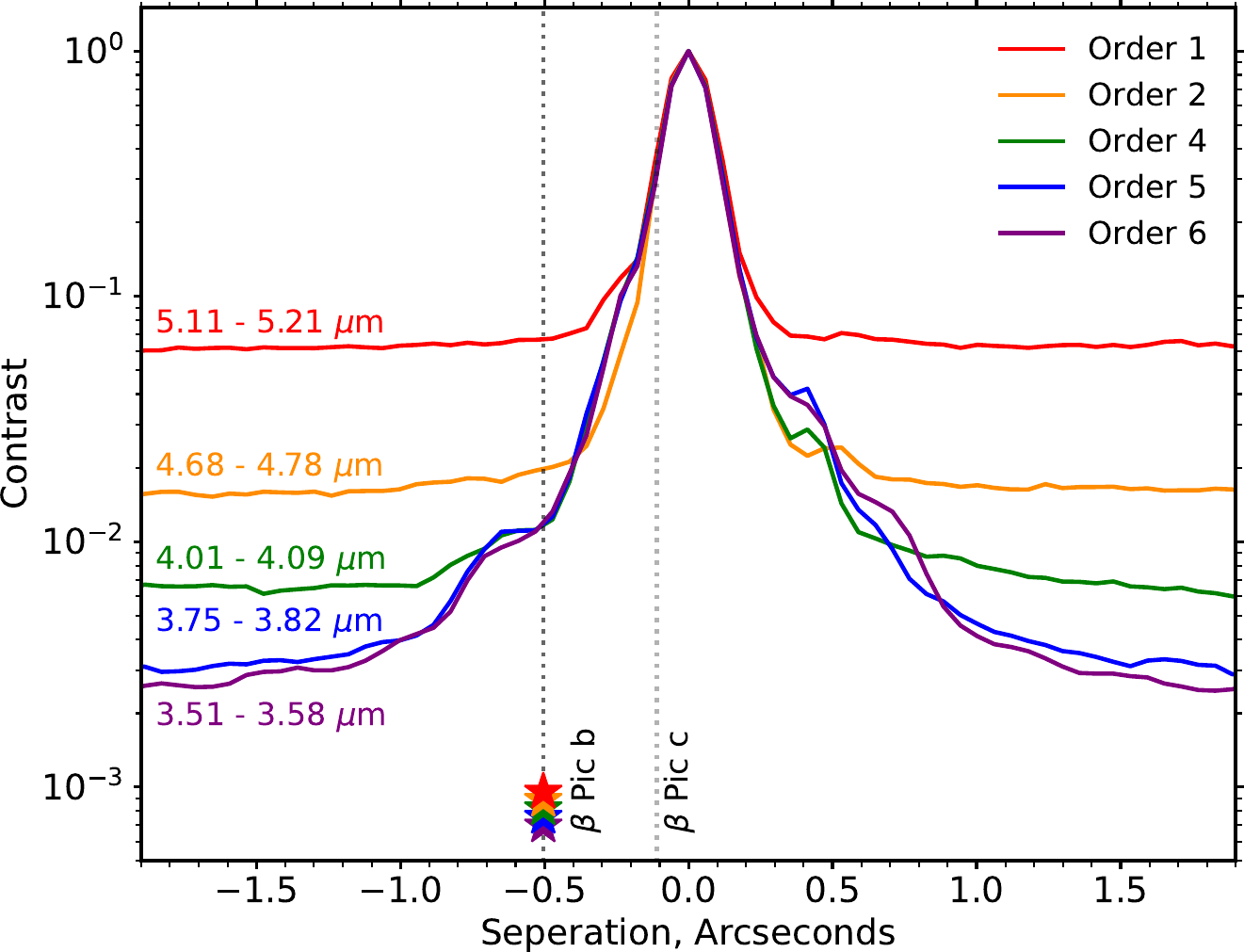}
    \caption{The spectrally averaged spatial profile of each order. The predicted blackbody contrast of $\beta$ Pic b for each order is denoted by a star at the planet spatial separation. In the first and second orders (red and orange) the stellar profile rapidly approaches a systematic noise floor at a separation of $\pm$~0.3~arcsec, driven by the thermal noise, and the spectrum at the planet position is background noise dominated. The forth, fifth, and sixth orders, however, do not reach the thermal noise floor until separations of >1\arcsec, beyond the planet position. The `shoulder' in the spatial profile at a separation of $\approx$~+0.5\arcsec is a result of the observed `secondary trace' in the data (see Section \ref{sec:Bad Pixels}). The spatial position of this phenomenon is wavelength dependent and appears consistent with diffraction effects.}
    \label{fig:PSF_contrast}
\end{figure}

\subsubsection{HRCCS in the background limited regime}
\label{sec:HRCCS in the background limited regime}

\begin{figure}
    \includegraphics[width=\columnwidth]{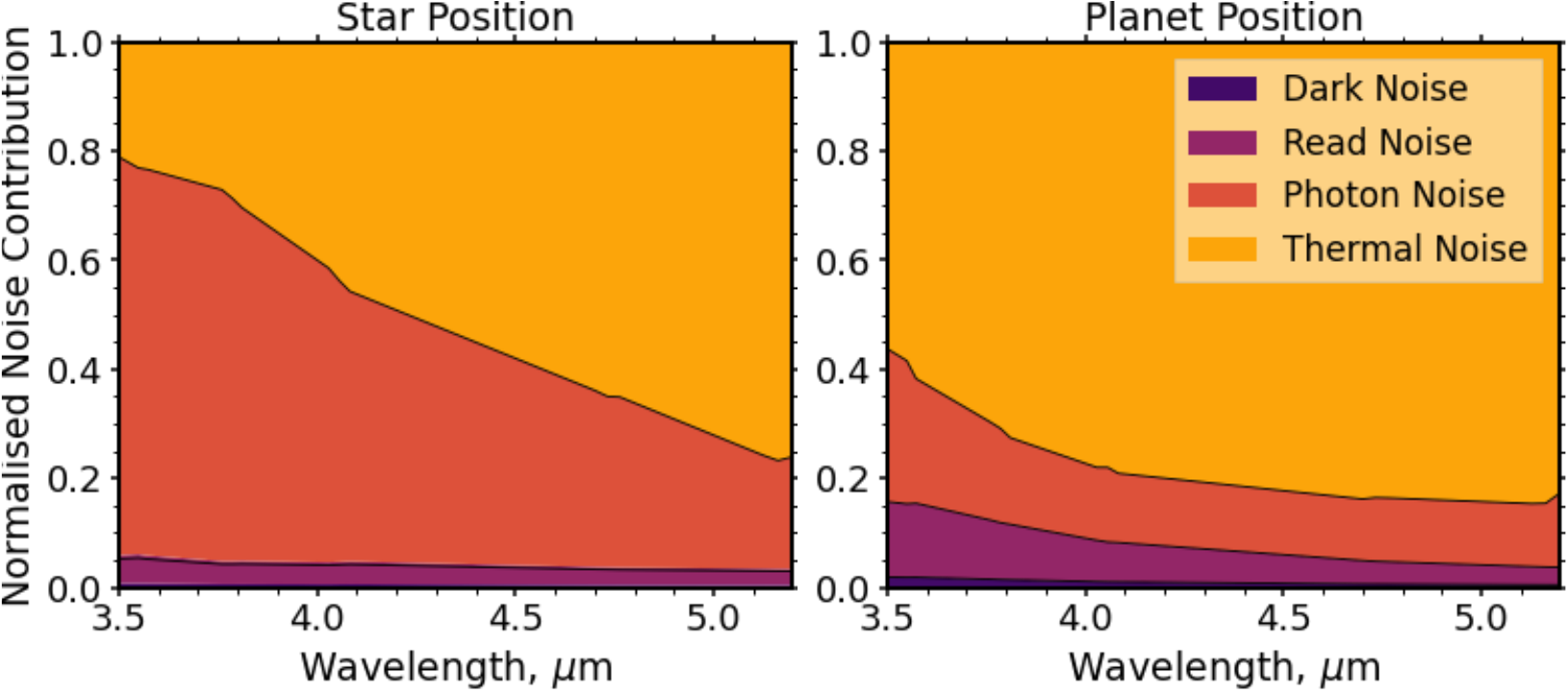}
    \caption{The normalised contribution of noise sources at the planet and star spatial positions for our CRIRES+ observations of $\beta$ Pic b, extrapolated from the non-continuous coverage of the M4368 grating setting. The sky thermal background noise dominates at the planet position for all orders of our data, from 3.51--5.21~$\mu$m. For classical Doppler shift HRCCS, where the signal is extracted at the stellar spatial position, the main noise source is photon noise up to $\approx$ 4~$\mu$m, but with a significant thermal noise component. Estimates of dark and read noise are taken from predictions of CRIRES+ performance, while the thermal noise and photon noise are measured directly from the \textit{M}-band data.} 
    \label{fig:Noise_stats_comb}
\end{figure}

The S/N for a planetary atmosphere detection using HRCCS, in an idealised case where every line has equal and complete absorption depth and the data are photon-noise dominated \citep{Snellen2015,Birkby2018}, is given by:

\begin{equation}
S/N_p = \left(\frac{F_p}{F_{*}}\right) \,~S/N_{*}~\sqrt{N_{\text{lines}}}
\end{equation}
\label{eq:3}

where $F_p$ is the flux from the planet, $F_{*}$ the flux from the star, $S/N_{*}$ is the total S/N of the stellar spectra and $N_{\text{lines}}$ is the number of detectable spectral lines of the target molecule. In the \textit{L} and \textit{M}-bands, even for bright targets such as the \textit{M$^{\prime}$} = 3.46~mag $\beta$ Pic A, the noise statistics are found to be increasingly dominated by thermal background noise. The increasing thermal background noise across the \textit{L} and \textit{M}-bands drives a rising noise floor with increasing wavelength (Fig.~\ref{fig:PSF_contrast}). The thermal noise contribution is the dominant source of noise across our entire observed wavelength range for the $\beta$ Pic CRIRES+ data at the planet position (Fig.~\ref{fig:Noise_stats_comb}) and we must consider a background noise component $\sigma_{\text{bg}}^2$, the wavelength averaged thermal background noise per exposure. Dark and read noise components are assumed to be negligible in these data. Additionally, in the case of HCS, the AO suppresses the stellar signal at the position of the planetary companion, expressed by the factor K \citep{Snellen2015}:

\begin{equation}
S/N_p = \frac{F_{\text{p}}}{\sqrt{F_{*}/K + \sigma_{\text{bg}}^2}} \,~\sqrt{N_{\text{lines}}}
\end{equation}
\label{eq:4}

We note that the planetary spin will also heavily impact the achieved S/N, as the rotational broadening of the planetary lines reduces the number of resolvable molecular lines in the data. When measured at the stellar position, the thermal noise only becomes the dominant noise source at wavelengths redder than 4.3~$\mu$m. This situation is more representative of classical HRCCS, in which the stellar trace is directly extracted as a 1D spectrum, and the Doppler shift from the planet orbital motion is used to disentangle the planetary signal from telluric and stellar signals. Therefore, it is suggested that classical HRCCS with CRIRES+ can operate in the photon dominated noise regime for the entirety of the \textit{L}-band, albeit with a non-negligible thermal noise component, and exclusively for very bright host stars.

\subsubsection{CRIRES+~/~VLT}
\label{sec:CRIRES+/VLT}

\begin{figure}
    \includegraphics[width=\columnwidth]{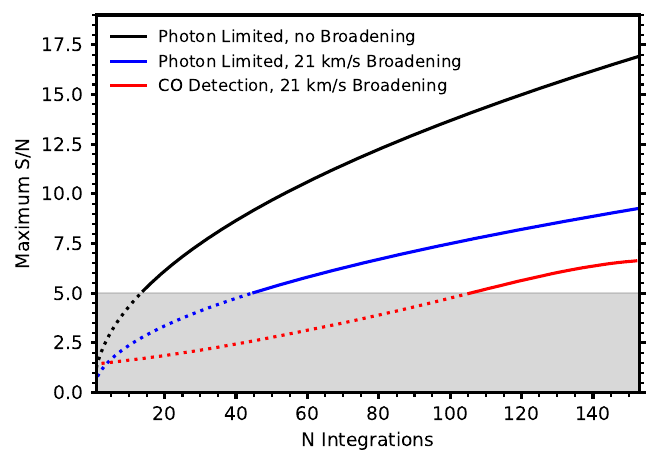}
    \caption{The dependence of the measured S/N of the CO detection in our data on the number of integrations included (red), constructed from stochastically combining the cross-correlation signals from the CO model at the instrument resolution across the 153 usable frames, averaged over 100,000 realisations. The shaded region masks signals with S/N <~5, the limit for a signal to be considered a robust detection. The 4.73~$\mu$m, thermal background limited, CRIRES+ data show a clear deviation from the idealised prediction of the photon limited case (black, \citealt{Birkby2018}) calculated using equation \ref{eq:3}, although a substantial fraction of the reduction in S/N can be attributed to the broadening of the planetary signal, which reduces the number of detectable spectral lines (blue, equation \ref{eq:3} using the broadened CO model).
    }
    \label{fig:SNR_vs_Nits}
\end{figure}

We have demonstrated that HRCCS in the \textit{M}-band using CRIRES+ is possible but challenging. Foremost among the challenges presented is the thermal background emission, which is the dominant noise source for HCS across the \textit{M}-band, and a sizeable component for classical HRCCS (Fig.~\ref{fig:Noise_stats_comb}). An additional problem posed by the thermal emission for CRIRES+ is the rapid saturation of the reddest orders in the \textit{M}-band by sky emission lines. This dictates the use of short exposure times, limited to a maximum of 45s in the \textit{L} and \textit{M}-bands for CRIRES+, but in practice to avoid saturation in the \textit{M}-band a maximum exposure length of 20s is required. The primary effect of this is to limit HRCCS observations to bright targets and necessitates a large number of integrations to accumulate sufficient S/N on source. The dependence of our retrieved S/N of the 4.73~$\mu$m CO detection (using the model broadened to the measured planetary spin-rotation) on the number of integrations used is compared to predictions from scaling relations in Fig.~\ref{fig:SNR_vs_Nits}. The thermal noise dominated \textit{M}-band observations show S/N improvements with increasing numbers of integrations used but require more observing time than is predicted in the idealised photon dominated case. The largest reduction in S/N is driven by the rotational broadening of the planet signal, which dramatically reduces the number of resolvable spectral lines (Fig.~\ref{fig:models}).

Additional problems include considerable telluric contamination, non-default wavelength calibrations, and complications modelling slit curvature effects, but these are surmountable using established HRCCS post-processing techniques and custom analysis routines. For future observations in the \textit{M}-band using CRIRES+ we recommend the following considerations. Foremost, we caution that the noise properties of high-resolution data dominated by thermal noise are more complex than the photon dominated environment in the previously explored NIR and visible bands. While in the photon-limited regime approximations are widely used for planning observations \citep{Birkby2018}, in the \textit{M}-band we strongly recommend end-to-end modelling of the detection significance to calculate the required exposure times, fully accounting for the thermal background. Finally, we reiterate that many of the default ESO calibration routines are not yet optimised for high-resolution studies in the \textit{M}-band and particular care must be taken in the extraction of spectra from the raw data. Alongside CRIRES+, for future studies in the \textit{M}-band, NIRSPEC/Keck in AO mode (NIRSPAO, R~$\approx$~50,000) could also feasibly be used for HRCCS, trading the spectral resolution of CRIRES+ for the increased aperture and photon collecting power of Keck \citep{McLean2000, WangJi2018}.

\subsubsection{METIS~/~ELT}
\label{sec:METIS/ELT}

The Mid-infrared ELT Imager and Spectrograph (METIS), a first-light instrument for ESO's ELT, will be revolutionary for HRCCS studies of exoplanets. METIS will be equipped with a suite of science modules, including an IFU-fed, diffraction-limited, R~=~100,000 spectrograph, operating in the \textit{L} and \textit{M}-bands \citep{Brandl2021}. This mode, specifically designed for HRCCS, offers a number of key advantages over CRIRES+ in the \textit{M}-band. First, the thermal noise and saturation risk of sky emission lines, which we identify as the greatest challenges in our data, are greatly reduced. The increased aperture of the ELT provides a reduction in the sky thermal background proportional to D$^4$, where D is the diameter of the primary mirror. Despite this remarkable reduction in background noise achievable with METIS, accurately modelling and reducing this noise component remains a key challenge \citep{Heikamp2014,Brandl2018}. The enhanced spatial resolution of the ELT permits a reduced Nyquist sampled plate scale: METIS will have a 0.0082~\arcsec/pix plate scale \citep{Brandl2021}, a factor of seven improvement on the 0.059~\arcsec/pix plate scale of CRIRES+. HCS can therefore be used to observe more closely separated companions, for example the planet $\beta$ Pic c, which in our data is only separated from the host star by two pixels and is therefore unrecoverable in the post-processing procedure. Finally, the increased aperture and photon collecting power of the ELT, with a reduced risk of the sky emission line saturation seen with CRIRES+/VLT, will facilitate observations of fainter targets in the \textit{M}-band than is possible with CRIRES+. 

Using the approximated expression for the signal-to-noise achieved by HCS with a thermal noise component (equation (\ref{eq:4})) we can estimate, to first order, the predicted S/N improvement of METIS compared to CRIRES+. This is performed by scaling our present understanding of CRIRES+/VLT in the \textit{M}-band for the ELT mirror size. We use as an example our CRIRES+ detection of CO at 4.73~$\mu$m. We scale the planet and stellar signals with aperture as D$^2$, and the background noise component with D$^{-2}$. The signal-to-noise of CRIRES+ on the ELT compared to the VLT is therefore approximated as:

\begin{equation}
\frac{S/N_{\text{ELT}}}{S/N_{\text{VLT}}} = \frac{D_{\text{ELT}}}{D_{\text{VLT}}}~\left( \frac{\frac{F_{\star}}{K} + \sigma_{\text{bg}}^2 \left(\frac{D_{\text{ELT}}}{D_{\text{VLT}}}\right)^{-4}}{\frac{F_{\star}}{K} + \sigma_{\text{bg}}^2} \right)^{-\frac{1}{2}}
\end{equation}

where $F_{\star}$ and $\sigma_{\text{bg}}^2$ are the CRIRES+ stellar signal and background noise measured in this work. Under the assumption that the only changes between METIS and CRIRES+ are the telescope mirror diameter, this calculation predicts a nominal S/N improvement factor of $\approx$7.5 between CRIRES+ and METIS for this specific science case. However, we caution that a myriad of parameters including, but not limited to, the wavelength coverage, relative throughput, instrument stability, the level of AO correction, achieved spectral resolution, instrument specific systematic effects, the quality of telluric correction, and the average conditions at the observing site will all substantially impact the relative S/N achieved with METIS.  Comprehensive end-to-end simulations, including all astrophysical and instrumental effects, are required to simulate HRCCS with ELT instrumentation (e.g. \citealt{Vaughan2024}). Here we have assumed an AO suppression factor, K, of 100 for both instruments but note that for METIS this suppression factor, and therefore signal-to-noise, will be reached at a separation of just 0.07\arcsec, a factor of seven improvement from CRIRES+ due to the improved ELT diffraction limit. The first order predicted S/N improvement relative to CRIRES+ suggests that METIS will achieve the same sensitivity as CRIRES+ for HCS in $<2\%$ of the exposure time, and that the S/N~=~6.6 CO detection achieved in this work could be obtained in as little as 30 seconds of integration time with METIS. For the most favourable targets, such an increase in precision will permit the search for exomoons using measurements of the planetary radial velocity \citep{Vanderburg2021,Ruffio2023} and monitoring of the time-dependent \textit{L} and \textit{M}-band spectral variability of spatially resolved planets over their full rotation periods \citep{Snellen2014, Sutlieff2023}. While we focus on METIS/ELT, we note that high-resolution \textit{M}-band spectroscopy will also be provided by GMTNIRS, a first-generation high-resolution spectrograph with R~=~85~000 in the \textit{L} and \textit{M}-bands, planned for the Giant Magellan Telescope (GMT) \citep{Lee2022}. The Thirty Meter Telescope (TMT) also has proposals for a second-generation high-resolution \textit{LMN}-band imager and IFU, MICHI \citep{Packham2018}.

\section{Conclusions}
\label{sec:Conclusions}

Using \textit{M}-band CRIRES+ observations of $\beta$ Pic b, we trial the use of high-resolution spectroscopy techniques in the unexplored 3.51--5.21~$\mu$m wavelength range. CO absorption is detected from the planet atmosphere at a maximum S/N of 6.6 at 4.73~$\mu$m, validating for the first time the functionality of HRCCS methods beyond 3.5~$\mu$m, in the regime in which the thermal background provides the dominant contribution to the noise. Alongside CO, the detection of H$_2$O (S/N~=~5.7) validates our \textit{M}-band measurements against benchmark studies detecting CO and H$_2$O in the \textit{K}-band. The marginal evidence of gaseous SiO (S/N~=~4.3) demonstrates the potential of HRCCS in this novel wavelength range to search for molecular species that are challenging to access at shorter wavelengths or lower spectral resolutions, but also the complexity associated with distinguishing genuine detections from correlated noise structures. We propose three scenarios which could result in observable abundances of SiO; the absence of magnesium-silicate clouds, in-homogeneous cloud coverage, or clouds with a particle size distribution that results in 4~$\mu$m transparency. Further data is required to confirm or refute the signal of SiO and to determine which, if any, of the cloud scenarios provides an appropriate description of the atmosphere of $\beta$ Pic b. The CO, H$_2$O, and SiO signals are rotationally broadened from the spin-rotation of $\beta$ Pic b, and a planetary rotation rate of $v$sin(i)~=~22$\pm$2 km~s$^{-1}$ is measured from the H$_2$O cross-correlation, consistent within error with previous studies in the \textit{K}-band. The highest detection significance, S/N~=~7.5, is achieved using a broadened model containing CO, H$_2$O, and SiO opacities, which represents our best approximation of the atmosphere of $\beta$ Pic b. Finally, we highlight the challenges in adapting HRCCS techniques to this novel wavelength range using existing instrumentation. The greatest observational obstacles are found to be the extreme telluric contamination, the strong thermal background noise, and the bright sky emission lines, which limit the feasibility of HRCCS techniques in the \textit{M}-band using CRIRES+ exclusively to targets with the brightest host stars and the most favourable contrast ratios. We have trialled post-processing techniques designed to approach these observational challenges, with the aim of gaining a practical first understanding of the behaviour of HRCCS across the 3--5~$\mu$m region. Advancements in instrumentation, most prominently the advent of METIS/ELT and corresponding instruments for the GMT and TMT, will revolutionise the use of HRCCS in the \textit{L} and \textit{M}-bands, permitting the search for biosignatures on nearby rocky exoplanets using HCS. The validation of high-resolution spectroscopy techniques in the thermal noise dominated \textit{M}-band with CRIRES+ is a crucial step to unlocking the full potential of this wavelength range with the ELTs.

\section*{Acknowledgements}
We thank the anonymous referee for their helpful and timely comments that improved the quality of the manuscript. LTP, JLB, LvS, SRV, and MEY acknowledge funding from the European Research Council (ERC) under the European Union’s Horizon 2020 research and innovation program under grant agreement No 805445. JLB further acknowledges the support of the Leverhulme Trust via the Philip Leverhulme Physics Prize. JPW acknowledges support from the Trottier Family Foundation via the Trottier Postdoctoral Fellowship. 

This research has made use of the NASA Exoplanet Archive, which is operated by the California Institute of Technology, under contract with the National Aeronautics and Space Administration under the Exoplanet Exploration Program. This research has made use of NASA’s Astrophysics Data System Bibliographic Services and the SIMBAD database, operated at CDS, Strasbourg, France. This research made use of SAOImageDS9, a tool for data visualization supported by the Chandra X-ray Science Center (CXC) and the High Energy Astrophysics Science Archive Center (HEASARC) with support from the JWST Mission office at the Space Telescope Science Institute for 3D visualization \citep{Joye2003}. This work made use of the whereistheplanet\footnote{\url{http://whereistheplanet.com/}} prediction tool \citep{Wang2021b}. This work has made use of the Python programming language\footnote{\url{https://www.python.org/}}, in particular packages including NumPy \citep{Harris2020}, SciPy \citep{Virtanen2020}, Matplotlib \citep{Hunter2007}, and Astropy \citep{Astropy2013,Astropy2018,Astropy2022}.

\section*{Data Availability}
The raw data used in this study is available for download from the ESO Data Archive under Programme ID 109.23G2.001. Processed data products and models are available on reasonable request to the corresponding author.



\bibliographystyle{mnras}
\bibliography{betapic}



\appendix
\label{sec:Appendix}

\section{Additional Figures}
\label{sec:App}

\begin{figure*}
    \begin{center}
    \includegraphics[width=1.98\columnwidth]{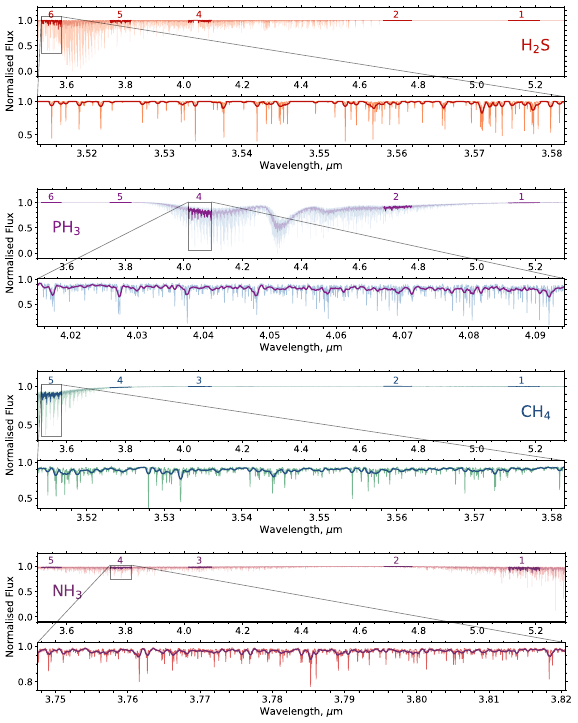}
    \caption{The normalised and continuum subtracted molecular models from \textit{petitRADTRANS} for the molecules which are undetected in this work. The models at the instrument resolution of R~=~100,000 are show in light colours, while the models broadened to the measured 22 km~s$^{-1}$ spin-rotation of $\beta$ Pic b (plotted in bold) demonstrate the reduction in both the number of lines and line depth due to rotational broadening. Highlighted regions of the spectra denote wavelengths covered by the five usable CRIRES+ M4368 detector orders. The detectability of these molecules is hampered by their lower predicted abundances, and a molecular structure with fewer well-spaced and deep lines in the CRIRES+ M4368 wavelength coverage.}
    \label{fig:all_other_models}
    \end{center}
\end{figure*}

\begin{figure*}
    \begin{center}
    \includegraphics[width=2\columnwidth]{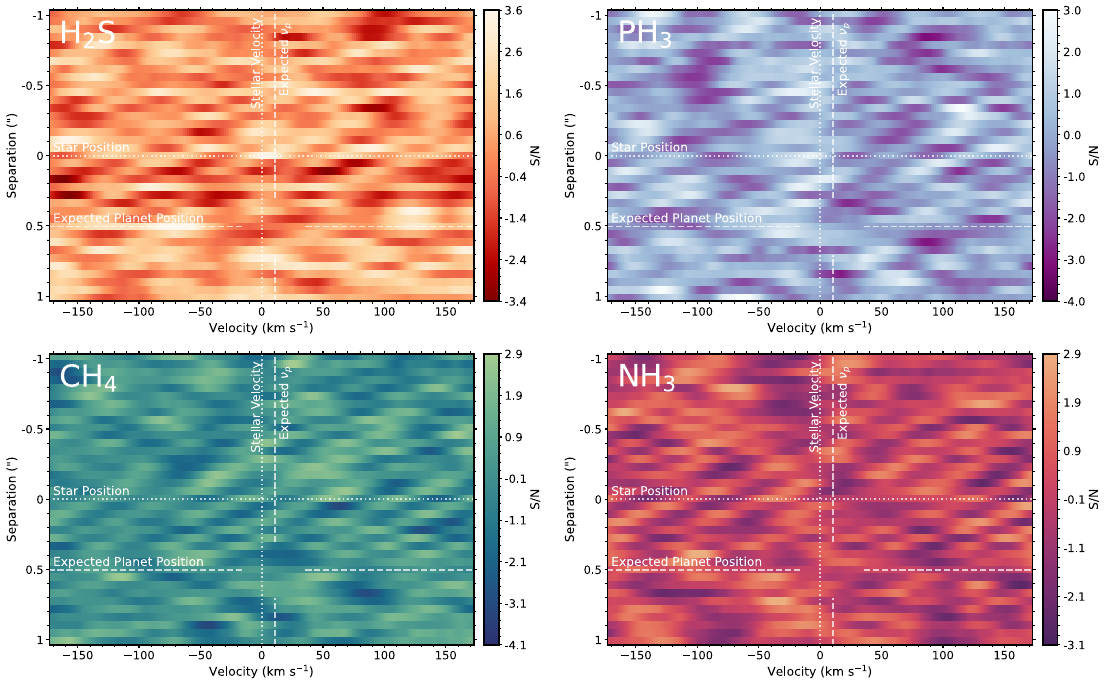}
    \caption{The cross-correlation S/N maps of $\beta$ Pic b for the undetected molecules in the \textit{M}-band (H$_2$S, PH$_3$, CH$_4$, and NH$_3$) using models that have been broadened to the measure rotational velocity of $\beta$ Pic b, as a function of spatial separation and velocity shift, centred at the systemic velocity of $\beta$ Pic A. Cross-correlation using models at the instrument spectral resolution likewise yield no peaks in the CCF.}
    \label{fig:all_other_ccfs}
    \end{center}
\end{figure*}

\begin{figure}
    \includegraphics[width=\columnwidth]{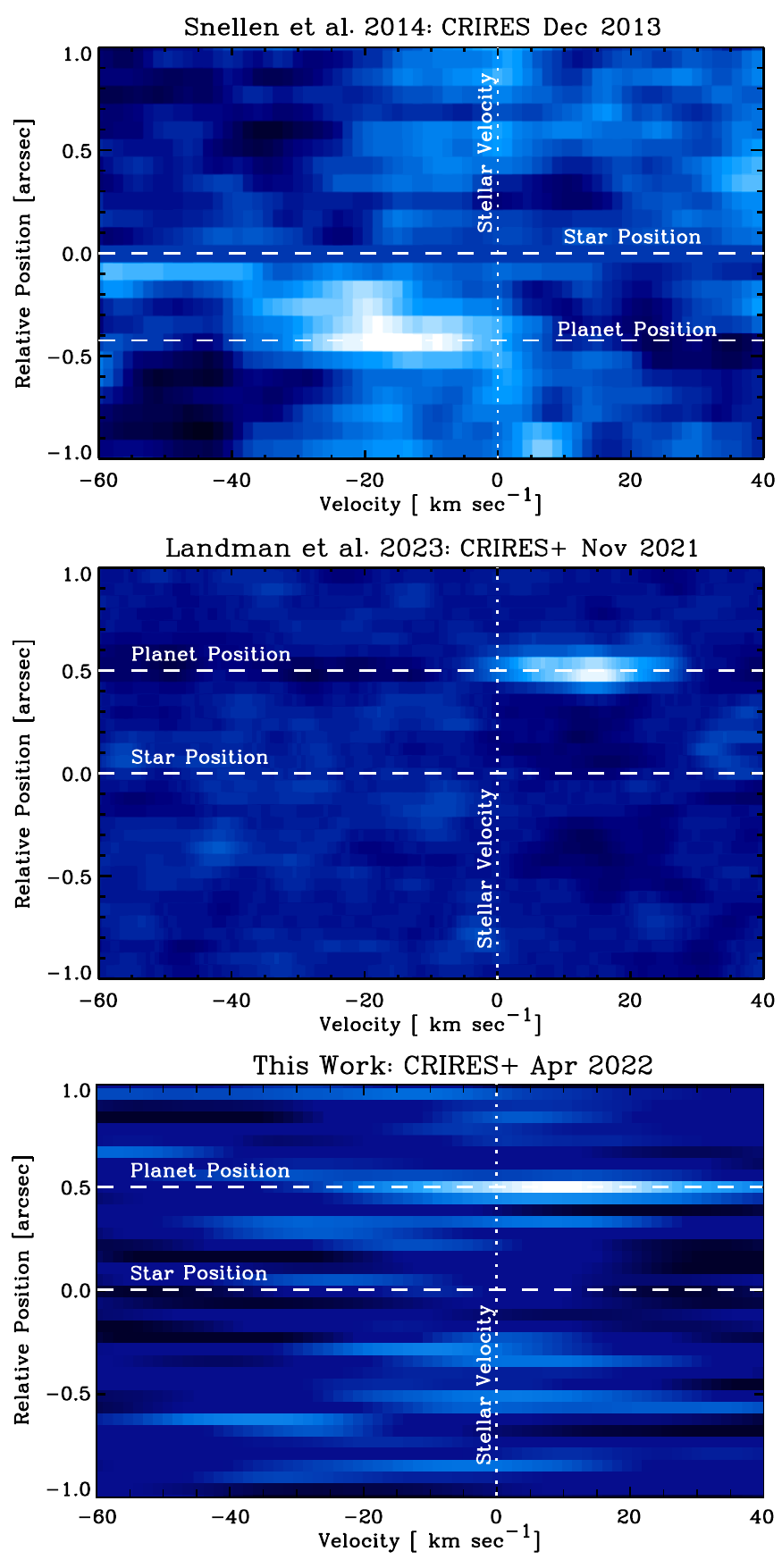}
    \caption{A visual comparison between the cross-correlation detection map of $\beta$ Pic b in the \textit{K}-band using the original CRIRES \citep{Snellen2014}, the recent CRIRES+ \textit{K}-band detection presented in \citep{Landman2023b}, and our \textit{M}-band CRIRES+ CO detection using broadened models. The \textit{M}-band CCF map is presented at the instrumental sampling of CRIRES+, consistent with the previous studies. Note that the CO detection presented in this work is visually broader as a result of the broader auto-correlation function of CO in the \textit{M}-band. The relative position along the slit and the colour map have been matched to the convention used in \citet{Snellen2014}. Figure adapted from \citet{Snellen2014} and \citet{Landman2023b}.
    }
    \label{fig:comparison}
\end{figure}

\begin{figure}
    \includegraphics[width=\columnwidth]{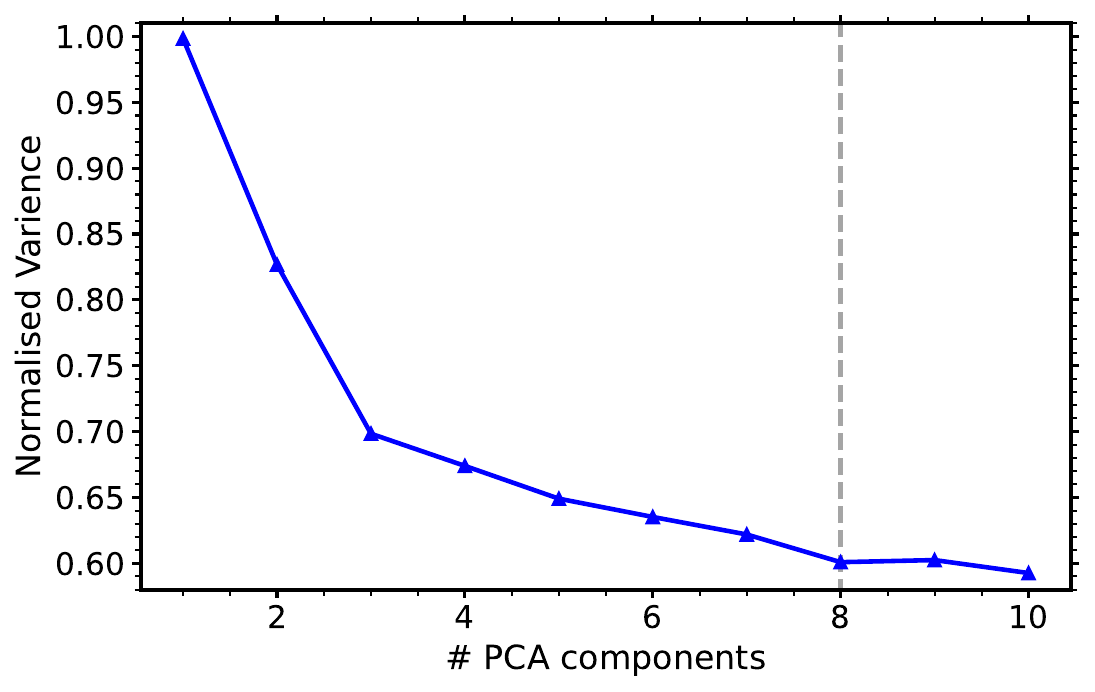}
    \caption{The impact of modifying the number of PCA components removed on the variance of the data. Initial reduction in the data variance is driven by the removal of telluric trends but beyond eight PCA components the data variance plateaus, suggesting that the planetary signal is eroded by further PCA iterations. We therefore select eight PCA components to be removed across all detector orders.}
    \label{fig:SNR_vs_NPVA}
\end{figure}

\begin{figure*}
    \centering
    \includegraphics[width=2\columnwidth]{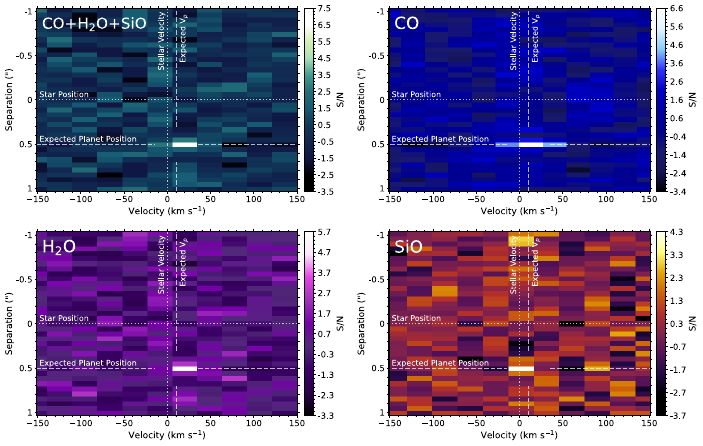}
    \caption{Cross-correlation maps produced on cross-correlation with molecular models broadened to the measured rotational broadening of each molecule, displayed at the resolution used to generate the estimate of the noise for Fig.~\ref{fig:All_broad}. The CCF is calculated at the 1.5~km~s$^{-1}$ instrument sampling, but is resampled every $\sqrt{2}$~$v_{\text{rot}}$ (shown here), where $v_{\text{rot}}$ is the measured rotational broadening for each molecule. The standard deviation of this down-sampled CCFs is then used as the estimate of the noise of each row, in order to reduce the impact of correlated signals introduced through the cross-correlation of the broadened model and the broadened planetary lines in the data.}
    \label{fig:All_broad_pixelated}
\end{figure*}


\bsp	
\label{lastpage}
\end{document}